\documentstyle[12pt]{article}
\input epsf

\newcommand{\mybibitem}[1]{\bibitem{#1}} 

\newcommand{\be}[1]{\begin{eqnarray} \label{#1} }
\newcommand{\eq}{\end{eqnarray}}

\def\bac{\begin{equation}\begin{array}{rll}}
\def\ea{\end{array}\end{equation}}
\def\vare{\varepsilon}

\def\au{{\underline a}}
\def\bu{{\underline b}}

\def\a{\alpha}
\def\b{\beta}

\def\d{\delta}
\def\e{\epsilon}        
\def\f{\phi} 

\def\k{\kappa} 
\def\l{\lambda}

\def\q{\theta} 
\def\t{\tau}

\def\G{\Gamma}

\def\L{\Lambda}

\def\ci{{\cal{I}}}

\def\car{{\cal{R}}}

\def\half{\frac{1}{2}}
\def\pr{^{\prime}}      
\def\prr{^{\prime\prime}}       

\newcommand{\NPB}[1]{Nucl.\ Phys.\ {\bf B#1}}
\newcommand{\PLB}[1]{Phys.\ Lett.\ {\bf B#1}}

\newcommand{\PRL}[1]{Phys.\ Rev.\ Lett.\ {\bf B#1}}
\newcommand{\PRD}[1]{Phys.\ Rev.\ {\bf D#1}}

\makeatletter
\@addtoreset{equation}{section}
\@addtoreset{equation}{subsection}
\def\theequation{\ifnum\value{section}=0 \arabic{equation}
\ignorespaces \else
\ifnum\value{subsection}=0 \thesection.\arabic{equation}\ignorespaces \else
\thesection.\arabic{subsection}.\arabic{equation}\ignorespaces \fi
\fi}
\@addtoreset{table}{section}
\@addtoreset{table}{subsection}
\def\thetable{\ifnum\value{section}=0 \arabic{table} \ignorespaces \else
\ifnum\value{subsection}=0 \thesection.\arabic{table}\ignorespaces \else
\thesection.\arabic{subsection}.\arabic{table}\ignorespaces \fi
\fi}
\makeatother

\newcommand{\bw}{{\bar W}}

\newcommand{\fb}{\bar {\f}}

\newcommand{\FF}{{\cal F}}
\newcommand{\HH}{{\cal H}}

\newcommand{\ad}{{\dot\alpha}}
\newcommand{\md}{{\dot\mu}}
\newcommand{\bd}{{\dot\beta}}
\newcommand{\del}{\partial}
\newcommand{\Del}{\nabla}
\newcommand{\Delb}{{\bar\nabla}}

\newcommand{\shalf}{\hbox{$\frac12$}} 
\newcommand{\squart}{\hbox{$\frac14$}} 
\newcommand{\ret}{\nonumber\\} 

\begin{document}

\begin{titlepage}

\begin{flushright}
ITP-SB-96-64 \\
USITP-96-14 \\
hep-th/9612195 \\
\end{flushright}

\begin{center}
\vskip3em
{\large\bf Monopoles in Quantum Corrected\\
\vskip .5em
N=2 Super Yang-Mills Theory}

\vskip3em
{G.\ Chalmers,\footnote{E-mail:chalmers@insti.physics.sunysb.edu}
M.\ Ro\v cek,\footnote{E-mail:rocek@insti.physics.sunysb.edu}}\\
\vskip .5em {\it Institute for Theoretical Physics\\
State University of New York\\
Stony Brook, NY 11794-3840, USA\\}
\vskip .5cm

{\normalsize R.\ von Unge\footnote{E-mail:unge@vanosf.physto.se}\\ \vskip
.5em{\it
Institute of Theoretical Physics \\
University of Stockholm, Box 6730 \\
S-113 85 Stockholm SWEDEN}\\
\vskip2em
}
\end{center}

\vfill

\begin{abstract}

\noindent
We study the low-energy effective Hamiltonian of $N=2$ super
Yang-Mills theories.  We find that the BPS equations are unchanged
outside a quantum core where higher dimension contributions are
expected to be important.  We find {\it two} quantum generalizations
of the BPS soliton.  The leading higher-derivative correction to the
effective action is shown not to contribute to the BPS mass formula.
\end{abstract}

\vfill
\end{titlepage}
\section{Introduction}\label{intro}

Over the last two years, there has been dramatic progress in our
understanding of the low-energy structure of supersymmetric gauge
theories: The complete low-energy effective action for $SU(2)$ super
Yang-Mills theory has been constructed \cite{sw}.

Classical monopole and dyon configurations have been of great interest
for many years.  However, little is known about their quantum
corrected form. In this paper, we assume that minima of the quantum
corrected Hamiltonian that follows from the low-energy effective
action \cite{sw} correctly describe quantum corrected BPS-saturated
states.  From this assumption, we derive a number of results: (1) The
BPS equations are {\em unmodified} by the quantum corrections outside
of a strongly-coupled quantum core that we cannot describe in
detail. Furthermore, we obtain the quantum corrected form of the BPS
mass formula. As one approaches the strongly coupled region, the
quantum core expands to fill all space and the solution breaks
down. (2) For spherically symmetric dyons ($n_m=\pm1$), our ignorance
of the detailed structure of the quantum core manifests itself as a
single parameter $\d$ in the asymptotic region. This parameter has a
physical significance: it describes the suppression of the massive
vector boson contribution to the soliton. (3) Deviations from the
classical magnetic field come entirely from $\d$. In contrast, the
electric field is no longer just a simple duality transform of the
magnetic field. (4) If we define a local central charge $Z(r)$, we
find {\em that its phase is constant over all space}.  This
allows us to compute (implicitly) the complete soliton solution
outside the quantum core, {\it e.g.}, to find the quantum corrected
electric field. (5) In particular, we find that the local
$\Theta$-angle can ``bleach'' as one approaches the quantum core.
(6) We find that depending on the asymptotic value of
the Higgs field $\f(\infty)$ at spatial infinity, there are two
different classes of solitons which differ in their boundary
conditions on the quantum core as well as in their properties under
analytic continuations of $\f(\infty)$. We explicitly
see how monodromies in the moduli space of vacua relate solitons in
different charge sectors. (7) We compute the contribution of a class
of higher dimension terms to the energy of a soliton, and find that
they vanish.

The outline of the paper is as follows. In section 2 we discuss
general aspects of the low-energy effective action of $N=2$
supersymmetric Yang-Mills theory.  In section 3 we find the BPS
equations and bound.  We next investigate spherically symmetric
solutions to the BPS equations. In section 5, we review the explicit
form of the low-energy effective action for the $SU(2)$ model and then
discuss the field equations from the effective action. We then discuss
the various quantum cores that can arise. In section 7 we discuss two
kinds of quantum-corrected solitons, and describe their features in
various representations. In section 8, we discuss the explicit spatial
dependence of the dyons, and the parameter $\d$ that encodes our
ignorance of the structure of the quantum core.  We then compute the
effect of a class of higher derivative corrections. In our
conclusions, we list a variety of natural extensions of our work. We
end with two appendicies.

\section{N=2 Low Energy Effective Action}

The leading term in a momentum expansion of the effective action
of $N=2$ super Yang-Mills theory is given by the imaginary part of
an $N=2$ chiral integral of a holomophic function $\FF(W)$\cite{FF}
\be{Fdef}
S_{\FF}={1\over 4\pi} {\ci}m\int d^4x\, d^4\q \FF(W) \ ,
\eq
where $W$ is the $N=2$ gauge superfield. In $N=1$ language this action is
\be{FinN=1}
S_{\FF}={1\over 4\pi} {\ci}m\left[\int d^4x\,
d^2\q\half\FF_{AB}(\f)W^{A\a}W^{B}_{\a}
+\int d^4x d^2\q d^2\bar{\q}\FF_{A}(\f)\bar{\f}^A \right] \ ,
\eq
where $\FF_A = (\partial/\partial{\f_A})\FF$.  The full
quantum moduli space on the Coulomb branch
has been described in \cite{sw} and will be
reviewed briefly below; the naive gauge-covariantization gives rise to
an action which reduces in the low-energy limit to the $U(1)$ theory.

We also consider the next term in the momentum expansion; it is given by a full
$N=2$ superspace integral of a real function $\HH(W,\bar{W})$ \cite{h,HH1} :
\be{Hdef}
S_\HH =\int d^4x\, d^8\q ~\HH(W,\bw) \ .
\eq
Written in $N=1$ superspace $S_\HH$ takes the form \cite{HH1}
\be{N1act}
S_\HH \!&=& \!\int d^4x \,d^2\q \,d^2 \bar{\q}\, \left(g_{A\bar
B}\left[-\shalf\Del^{\a\ad}\f^A\Del_{\a\ad}\fb^B +i\bar W^{B\ad} (\Del^\a{}_\ad
W^A_\a +\Gamma^A_{CD}\Del^\a{}_\ad\f^CW^D_\a)\right.\right.\ret
&&\qquad\qquad\qquad\qquad~~~
\raisebox{0.0em}[1.em][.9em]{$-(f^A_{CD}\,\bar W^{B\ad}\f^C\Delb_\ad\fb^D +
f^B_{CD}\,W^{A\a}\fb^C\Del_\a\f^D)$}\\ &&\qquad\qquad\qquad\qquad~~~\left.
+(\Del^2\f^B+\shalf\G^{\bar B}_{\bar C\bar D}\bar W^{C\ad}\bar W^D_\ad)
(\Delb^2\fb^A+\shalf\G^A_{EF}W^{E\a}W^F_\a)\right] \nonumber\\
&&\qquad\qquad\qquad~~\left. +\squart R_{A\bar BC\bar D}(W^{A\a}W^C_\a\bar
W^{B\ad}\bar W^D_\ad) -i\mu_A(\shalf\Del^\a W^A_\alpha -f^A_{BC}\f^B\fb^C)
\right)\,\, . \nonumber \eq
The functions $g,\G,R$ are defined in terms of the
partial derivatives of $\HH$
\be{geodefs} g_{A\bar B}=\HH_{A\bar B}\ ,
\qquad\Gamma^A_{BC}=g^{A\bar D}\HH_{BC\bar D}\ ,\qquad R_{A\bar BC\bar
D}=\HH_{AC\bar B\bar D} -g_{E\bar F}\Gamma^E_{AC}\Gamma^{\bar F}_{\bar B\bar D}\
, \eq and $\mu$ is the moment map defined by
\be{mom}
f^C_{AB}\HH_CW^B=\eta_A(W)+i\mu_A(W,\bar W)\ ,
\eq
for $\eta$ an arbitrary holomorphic function of $W$.  The expansion of
this $N=2$ term in the low-energy Lagrangian contains many terms and is
given in the appendix.  

\section{BPS Equations and Bound}

We first discuss the field equations and BPS states appearing in
the $N=2$ quantum corrected theory.  We consider configurations
without potential energy, {\it i.e.}, with $[\f, {\bar\f}]=0$,
throughout the paper.  The bosonic part of the action is then
\be{Ffunction}
S_\FF = -{1\over 4\pi} {\ci}m \int d^4x ~\FF_{AB} \Bigl\{ {1\over 2}
(B_i^A+iE_i^A)(B_i^B+iE_i^B) + \nabla_\au \f^A \nabla^\au
{\bar\f}_A \Bigr\} \ ,
\eq
where $\{\nabla_\au\}=\{\nabla_{\a\ad}\}$ with $\{\au\}=\{0,i\}$, and
$B_i^A=\half \epsilon_{ijk} F^{jk}$, $E_i^A=F_{0i}^A$.  For the
moment we ignore the higher derivative terms obtained from $S_\HH$; 
from $S_\FF$ we obtain the conjugate momenta
\be{pi}
\Pi_{iA} &\equiv & {\delta L\over \delta(\partial_0 A_i^A)}
  = -{1\over 4\pi} {\car}e \bigl\{ \FF_{AB} ( B_i^B +i E_i^B) \bigr\}
\nonumber\\ \\
\Pi_A &\equiv & {\delta L\over \delta(\partial_0 \f^A)}
 = -{1\over 4\pi} {\ci}m(\FF_{AB}) \nabla_0 {\bar\f}^B \ ,\nonumber
\eq
and the Hamiltonian
\be{HamDef}
H = \int d^3x ~\Pi_{iA} (\partial_0 A_i^A) + \Pi_A (\partial_0 \f^A)
+ {\bar\Pi}_A (\partial_0 {\bar\f}^A) - L \ .
\eq
Classically, the superpotential is $\FF=\half \tau\f^2$ with
$\tau={4\pi i\over g^2} +{\theta\over 2\pi}$.

The field equations are quantum corrected, whereas the Bianchi identities
are not; in particular, the quantum corrected form of the Gauss constraint is
\be{QantGauss}
\nabla_i \Pi_{iA} = -{1\over 4\pi} {\ci}m (\FF_{CD}) \epsilon^C_{AB}
\bigl( \f^B \nabla_0 {\bar\f}^D + {\bar\f}^B \nabla_0 \f^D \bigr) \ ;
\eq
This must be satisfied by the quantum-corrected soliton field
configurations.  We
consider static configurations and choose the gauge $\nabla_0 \f^A=0$; then
$\nabla_i \Pi_{iA} = 0$.  Note that classically the constraint
becomes
\be{ClassGauss}
\nabla_i \Pi_i^{A,cl} = -{1\over g^2} \nabla_i E_i^A + {\theta\over 8\pi^2}
 \nabla_i B_i^A =0 \ ,
\eq
which implies $\nabla_i E_i^A=0$ through the Bianchi identity.  In the quantum
theory, however, $\FF_{AB}$ is field dependent, and the electric and magnetic
fields may be regarded as living in a quantum corrected dielectric medium:
we have $\nabla_i\Pi_i^A=0$ with $\nabla_i E_i^A\neq 0$.  As we shall
see below, the quantum corrected Gauss constraint implies nontrivial
features of the $\Theta$-angle structure of the soliton solutions.

The bosonic part of the Hamiltonian (with $\nabla_0 \f^A=0$) is explicitly
\be{Fcomponent}
H= {1\over 8\pi}  {\ci}m \int d^3x \FF_{AB} \left( E_i^A E_i^B + B_i^A B_i^B
 + 2\Del_i \f^{A}\Del_i \bar{\f}^{B}\right) \ .
\eq
As usual, the BPS bound and first order form of the field equations are found
by writing the energy density as a squared function together with a boundary
term that gives the topological mass of the field configuration.  For
the quantum corrected theory we have $H=H_{\rm top}+H_0$ with
\be{almostBPS}
H_0 = {1\over 8\pi} {\ci}m \int d^3x \FF_{AB} \Bigl\{
 (B^A_i +iE^A_i+\sqrt{2} e^{i\alpha} \nabla_i\f^A)
 (B^B_i-iE^B_i+\sqrt{2} e^{-i\alpha} \nabla_i {\bar\f}^B) \Bigl\}  \ .
\eq
A constant phase $e^{i\alpha}$ has been introduced and will be given
explicitly below; it might appear that $e^{i\alpha}$ could be absorbed
into the complex field $\f$.  However, the asymptotic phase of $\f$
has an independent significance: it determines the $\Theta$-angle of the
vacuum.

In the form (\ref{almostBPS}), the Hamiltonian is written as a
positive quantity whenever ${\ci}m \FF_{AB}\geq 0$; the BPS equations
for the general monopole and dyonic states is
\be{BPS}
B_j^{B}+iE_j^{B}+e^{i\a} \sqrt{2} \Del_j\f^{B} =0  \ .
\eq
These equations have the same form as in the classical
Yang-Mills-Higgs action \cite{BPS,JZ}!\footnote{For the
(electrically-neutral) monopole, Seiberg and Witten \cite{sw} observed
that $H$ had the form of a square of BPS-type equations up to the mass
boundary term; they did not comment on the fact that the equations are
unmodified by quantum corrections, though it is clear from their
discussion.}

The energy associated with the boundary terms is found to be
\be{mass}
H_{top} = -\sqrt{2} \int d{\vec\Sigma} \cdot ~\Bigl\{ {\vec\Pi}_A
{\ci}m ( e^{i\alpha} \f^A) + {1\over 4\pi} {\vec B}^A {\ci}m
( e^{i\alpha} \FF_A ) \Bigr\} \ .
\eq
The quantization condition on the fields applies to $B_i$ and 
$\Pi_i$.  The asymptotic value of $B_i$ gives the homotopy class 
of the gauge field, and that of $\Pi_i$ is quantized through the 
invariance of $2\pi$ rotations around ${\hat\f}$ at spatial 
infinity \cite{wd}.  The asymptotic values of the scalar fields 
$\f(r)$ and $\f_D(r)$ are defined as
\be{asymptotic}
{\rm lim}_{r\rightarrow\infty} \f(\vec{r}) \equiv a(u) \quad\quad
{\rm lim}_{r\rightarrow\infty} \f_D(\vec{r}) \equiv a_D(u) \ ,
\eq
which are the vacuum expectation values of the scalar and its dual,
$a(u)$ and $a_D(u)$.  Their values are controlled in the low-energy 
theory by the complex order parameter $u$.  The electric and magnetic 
quantum numbers $n_e$ and $n_m$ are given by
\be{quant}
n_m a_D =-{1\over 4\pi} \int d{\vec\Sigma}\cdot {\vec B}^A \f^{D}_A\ ,
\quad\quad
n_e a = -\int d{\vec\Sigma} \cdot {\vec\Pi}_A \f^A  \ ;
\eq
the electric charge is, as always, given by an integral of $\vec E$, and for
$\Theta\neq 0$ does {\em not} vanish even when $n_e=0$ \cite{wd}.
In general the mass formula is saturated for dyonic states upon taking the
angle $\alpha$ to satisfy
\be{alpha}
e^{i\alpha}= i \frac{\bar Z}{\vert Z\vert} \quad\quad Z=n_e a+n_m a_D \ ,
\eq
where $Z$ is the central charge of the $N=2$ superalgebra. In this case we
obtain the BPS bound for the total energy
\be{bound}
E \geq \sqrt{2} \vert Z\vert \ .
\eq
As noted above, when the quantum corrected monopole/dyon solutions
satisfy the {\em  usual} BPS equations, the bound is saturated.

The field equations derived from the Lagrangian (\ref{Ffunction}) must
also satisfy the BPS equations.  Both sets of equations demand that
the fields and their duals satisfy their corresponding equations of
motion.

By differentiating the BPS equations (\ref{BPS}) and
imposing the Bianchi identities $\nabla_i B_i^A =0$ we find the
real part of the classical field equation for the scalar $e^{i\a}\f$
\be{scalarbox}
{\car}e (e^{i\alpha} \nabla_i\nabla_i \f^A) = 0 .
\eq
Inserting the BPS equations into the Gauss constraint we find the real
part of the quantum corrected field equation for $\f$, which is the
real part of the classical field equation for the dual scalar
($e^{i\a}\f\rightarrow e^{i\a} \f_D$)
\be{dualbox}
\nabla_i \Pi_{i,A} = {\car}e (e^{i\alpha} \nabla_i\nabla_i \f^D_A) = 0 \ .
\eq
Both equations need to be satisfied for the soliton solutions.  Classically,
the Gauss constraint enforces $\nabla_i E_i^A=0$; the complex
scalars in this case obey $\nabla_i\nabla_i\f^A=0$ and
$\nabla_i\nabla_i{\bar\f}^A=0$.  Since in the classical theory
$\f_D=\tau_{cl} \f$, both equations (\ref{scalarbox}) and
(\ref{dualbox}) are indeed satisfied simultaneously.

\section{Radial Ansatz}

In this section we discuss the quantum corrected solutions to the soliton
field equations.  We limit
ourselves to spherically symmetric field configurations for simplicity
and comment further on the general case when 
necessary.

The fields in a radial ansatz take the form
\be{radans}
\f^A = e^A \f(r)\ , \quad\quad A_i^A = \epsilon^A_{~ij} e^j
\left({1-L(r)\over r}\right)\ ,
\quad\quad A_0^A = e^A b(r) \ ,
\eq
where $e^A={\hat r}^A$ is a unit radial vector and $\f(r)$ is
complex.  In \cite{wg} it was shown that $n_m=\pm 1$ for spherically 
symmetric dyons.  We also note that just as in the classical theory, the
fundamental spherically symmetric monopole may be embedded in higher 
rank gauge groups through an $SU(2)$ subgroup \cite{embed}.
The potentials in (\ref{radans}) give rise to magnetic and electric
fields 
\be{berad}
B_i^A = e_i e^A {L^2-1\over r^2} + \Pi_i^A\, {L_r\over r}\ , \quad\quad
 E_i^A = -e_i e^A b_r - \Pi_i^A\, {bL\over r}\ ,
\eq
where we have defined the projector $\Pi_{AB}=\delta_{AB}-e_A e_B$.  
The radial vector $e^A$ is orthogonal to the $SU(2)/U(1)$ projector
$\Pi_{AB}$. 
The BPS equations (\ref{BPS}) for the spherically symmetric fields are
\be{radialBPS}
\sqrt{2} e^{i\alpha} \f_r = {1-L^2\over r^2} + ib_r
\quad\quad
\sqrt{2} e^{i\alpha} \f = -\frac d{dr}(\ln{L}) + ib  \ .
\eq
With this ansatz, $\sqrt{2}{\car}e \bigl( e^{i\a}\f\bigr)=-\frac
d{dr}(\ln{L})$ and the field equation (\ref{scalarbox}) is trivially
satisfied.  We define
\be{xdef}
X={\car}e (e^{i\alpha} \f)\ , \quad\quad
X_D={\car}e (e^{i\alpha}\f_D) \ .
\eq
The equations (\ref{radialBPS}) are valid in the presence of the quantum
corrections;  however, the solutions differ from the classical ones
since the {\em quantum} relation between $\f$ and $\f_D$ is nonlinear.

We now turn to finding the explicit solution to (\ref{radialBPS}). The
real part implies:
\be{Leq}
\frac{d^2}{dr^2}\left(\ln{L}\right) = {L^2-1\over r^2}\ .
\eq
The most general solution to this second order equation (which, after
a change of variables $L=re^\rho$, can be recast as a 1-dimensional Liouville
equation on the real half-line) involves two integration constants:
\be{generalL}
L={\kappa r\over \sinh[\kappa(r+\d)]} \ .
\eq
In the classical limit we impose the boundary condition that the field
configurations are regular everywhere; this implies that
$\d=0$. As we shall see, we have no reason to impose such a condition
in the quantum case.

Having solved for $L$, we can use (\ref{radialBPS}) to find $X$
(\ref{xdef}):
\be{xsol}
X=-\frac{1}{\sqrt2}\,\frac{d}{dr}\ln{L}=
\frac{\k}{\sqrt2}\left(\coth{[\k(r+\d)]}-\frac1{\k r}\right)\ .
\eq
In both the quantum and the classical case, the
constant $\k$ is determined by the asymptotic properties at $r=\infty$:
\be{Xbound}
X(\infty)={\car}e\bigl(e^{i\a}a(u)\bigr) = {\car}e\left(ia\frac{n_m \bar a_D+n_e
\bar a}{\vert n_m a_D + n_e a\vert}\right) = -{\ci}m\left(a\frac{n_m
\bar a_D+n_e \bar a}{\vert n_m a_D + n_e a\vert}\right)\ ,
\eq
where we have used the expression (\ref{alpha}) for $e^{i\a}$.
Expanding (\ref{xsol}), we find
\be{ksol}
\k =\sqrt{2}X(\infty)=\sqrt{2}n_m\frac{a\bar
a}{\vert n_m a_D + n_e a\vert} {\ci}m\left(\frac{a_D}a\right)\ .
\eq

Note that the solution (\ref{xsol}) satisfies the field
equation (\ref{scalarbox}), which, in the radial ansatz, becomes:
\be{o2eq}
X_{rr} + 2{X_r\over r} - {{L^2 X}\over r^2} = 0 \ .
\eq
Clearly, the dual field $X_D$ obeys the same equation
(see \ref{scalarbox} and \ref{dualbox}). The most general solution to 
equation (\ref{o2eq}) is an arbitrary linear combination of two 
basic solutions:
\be{twosol}
\Psi^{(1)} = \coth[\kappa(r+\d)]-{1\over \kappa r}\ , \quad\quad
\Psi^{(2)} = {1\over \kappa r} \coth[\kappa(r+\d)] \ .
\eq
The boundary conditions at $r=\infty$ are enough to determine $X_D$ in
terms of $X$: the constant term follows from the observation that
\be{reeiaz}
{\car}e(e^{i\a}Z)=n_mX_D(\infty)+n_eX(\infty)\ .
\eq
However, recall that $e^{i\a}=i\bar Z/\vert Z\vert$ (\ref{alpha});
this implies that ${\car}e(e^{i\a}Z)=0$, and hence
$X_D(\infty)=-(n_e/n_m)X(\infty)$. Furthermore, we can determine the
coefficients of the $1/r$ terms in $X$ and $X_D$. From (\ref{quant})
we find
\be{boundary}
B_i^A &=& -{e_i e^An_m\over r^2}
+ \ldots \nonumber\\ \\
\Pi_i^A &=& -{e_i e^A n_e\over 4\pi r^2} + \ldots \nonumber
\eq
On the other hand, the BPS-equation (\ref{BPS}) implies
\be{bpsimp}
B_i^A &=& - \sqrt{2} \nabla_i (e^A X) \ ,
\nonumber\\ \\
\Pi_{iA} &=& -{1\over 4\pi} \sqrt{2}\nabla_i (e_A X_D)\ .\nonumber
\eq
Expanding $X,X_D$ to order $1/r$, we find that the coefficient of the
$1/r$-term in $X_D$ is $n_e$.  This implies a simple important result:
\be{xxd}
n_mX_D(r)+n_eX(r)\equiv 0\ .
\eq
In other words, if we define a {\em local} central charge
$Z(r)=n_m\f_D(r)+n_e\f(r)$, then the phase of this central charge
is constant in position space $\vec{r}$ {\em even for the quantum 
corrected solution}:
\be{rez}
{\car}e\left(e^{i\a}Z(r)\right)=0\ .
\eq
In the classical case, the phases of $\f(r)$ and $\f_D(r)$ themselves
are constant; in addition, $\f_D(r)=\tau\f(r)$, and hence $\f(r)$ is
simply proportional to (\ref{xsol}) with $\d=0$ \cite{BPS,JZ}.

\section{Review of $SU(2)$ N=2 Yang-Mills theory}

The quantum corrected solutions we discuss arise in the low-energy
$N=2$ super Yang-Mills theory; in their seminal paper, Seiberg and
Witten \cite{sw} found an exact nonperturbative expression for the
$N=2$ superpotential\footnote{$\FF$ is sometimes called the
prepotential; as this term has been used for many years to refer to
the unconstrained superfields that arise in the solution of
constraints in superspace (see, e.g. \cite{book}), and as $\FF$ is the
$N=2$ analog of the $N=1$ superpotential, we refer to it as the $N=2$
superpotential.} $\FF$.  Holomorphy and monodromies of $\FF$ under duality
transformations imply that the moduli space of vacua in the low-energy
theory can be described in terms of period integrals of an underlying
torus.  The theory is described by the two functions $a(u)$ and
$a_D(u)$ (\ref{asymptotic}); these are sections of an $SP(2,Z)$ vector
bundle over a torus, and depend on the complex parameter $u$ that
labels the points on the Coulomb branch.

The scalar vacuum expectation values $a(u)$ and their duals $a_D(u)$ are
parameterized as integrals of a one-form $d\lambda$ around the two
homology 
cycles $\alpha$ and $\beta$ of a torus.  The gauge invariant complex
parameter $u$ (classically $u={\rm Tr}\f^2$) labels the inequivalent
vacua of the low-energy theory (or alternatively the complex 
structure of the underlying torus).  The relevant torus is determined by
the elliptic curve
\be{elcurcve}
y^2 = (x^2-1)(x-u) \ .
\eq
The vacuum expectation values are given as the integrals
\be{periods}
a(u)= \oint_\a d\l(u) =2\int_{-1}^1d\l(u)\ , \quad\quad
a_D(u) = \oint_\b d\l(u)=2\int_{1}^ud\l(u)\ 
\eq
where the meromorphic one-form is
\be{mero}
d\lambda(u) = {\sqrt{2}\over 2\pi} \Bigl({x-u\over x^2-1}\Bigr)^{1/2} dx \ .
\eq
The periods are explicitly computed from (\ref{periods}) to be
\be{shortint}
a(u) = {4\over\pi q}  E(q) \quad\quad q^2={2\over 1+u}
\eq
and
\be{longint}
a_D(u) = - i {4\over\pi q} \Bigl\{ E(q') -  K(q') \Bigr\} \ ,
\eq 
where $K(q)$ and $E(q)$ are complete elliptic integrals of the first
and second kind, respectively, and $q'=\sqrt{1-q^2}$ is the
complementary modulus.  We define the integrals by analytic
continuation from real $u>1$.

The superpotential and $U(1)$ coupling are
found from the period integrals:
\be{adandtau}
a_D (u) ={\partial\over\partial a} \FF (u)
\quad\quad
\tau (u) = {\partial a_D \over \partial a} = {\partial a_D\over\partial u}
\left({\partial a\over \partial u}\right)^{-1} \ .
\eq
Explicitly, we find the strikingly simple result
\be{exptau}
\tau(u)=i\frac{K(q')}{K(q)}\ .
\eq
The $S$-duality transformation $\tau\leftrightarrow-1/\tau$ 
interchanges the $\a$ and $\b$ cycles, that is, $q\leftrightarrow
q'$. 

We think of the soliton solutions described above as arising
from a spatially dependent $u({\vec r})$. In principal, $\f$ could
depend on other degrees of freedom, but we believe that this
can be ignored in regions where the low-energy effective action 
described by $\FF$ gives a good description of the physics.
The field configurations $\f({\vec r})$ and $\f_D({\vec r})$ are
determined through the period integrals:
\be{udependence}
\f({\vec r}) = \oint_\alpha d\lambda[u({\vec r})] \quad\quad
\f_D({\vec r}) = \oint_\beta d\lambda[u({\vec r})] \ .
\eq
Given a solution to the BPS equations it is a matter of inverting the
relations (\ref{udependence}) to find $u({\vec r})$.  Similar
constructions apply to solitons in higher-rank gauge groups G; in these 
cases, however, the moduli space is controlled by rank(G) complex 
parameters $u^i(\vec r)$ which must be solved for.

The low-energy theory has a duality group $SL(2,Z)$ that acts linearly on
$a_D(u)$ and $a(u)$.  A nontrivial subgroup of the duality group is
generated by monodromies around certain singularities in the theory.
The singularities at finite $u$ arise when solitonic states become
massless and the low-energy description in terms of the original
fields of the theory breaks down.  Explicitly, there
are monodromies around the points $u=\pm 1$ and $u=\infty$ that
generate the group $\Gamma(2)\subset SL(2,Z)$ and are given by
\be{mono}
M_1 = \pmatrix{ 1 & 0 \cr -2 & 1} \quad\quad
 M_{-1} = \pmatrix{ -1 & 2 \cr -2 & 3} \quad\quad
M_\infty = M_1~M_{-1} = \pmatrix{ -1 & 2 \cr 0 & -1} \ .
\eq
The mass formula is $SL(2,Z)$-invariant if we transform $(n_m,n_e)$
with the inverse of the matrix that transforms $(a_D,a)$:
\be{qnums}
Z\rightarrow \left(\pmatrix{ n_m & n_e }M^{-1}\right)\cdot 
\left( M \pmatrix{ a_D(u_i) \cr a(u_i) }\right) = Z \ .
\eq
The quantum numbers of the state that become massless at the point
$u_i$ are then determined by the left-eigenvector of $M_i$.
The monodromies in (\ref{mono}) occur around two strong
coupling singularities in the quantum moduli space at the points
$u=\pm 1$.  At these points
\be{singdef}
a_D(u=1) =0 \quad\quad  a(u=-1)-a_D(u=-1) =0  \ ,
\eq
so that by the BPS mass formula a $(1,0)$ monopole and $(1,-1)$ dyon becomes
massless.  These points appear in the low-energy theory as singularities in the
superpotential $\FF$.

Under a monodromy transformation in the semi-classical regime, the
magnetic and electric quantum numbers change according to
\be{monod}
\pmatrix{ n_m' & n_e' } = \pmatrix{ n_m & n_e } \pmatrix{ -1 & 2 \cr 0 & -1}
 = \pmatrix{ -n_m & 2n_m-n_e }
\eq
The semi-classical spectrum of stable bosonic states include the W-bosons
$(0,1)$ and fundamental dyons $(1,n_e)$ together with their conjugates.  Higher
magnetically charged states are neutrally stable and decay.

The soliton solutions presented in the following sections change 
accordingly under the transformations (\ref{mono}) but still satisfy 
their equations of motion.  We will present a simple interpretation 
of the action of $M_\infty$ on the solutions. 

\section{Soft core and Hard core}

In this section we discuss the range of validity of the BPS
equations (\ref{BPS}) and of the field solutions
derived from the low-energy effective action. There are two
features of the quantum corrected theory that have no
analog classically.

First, in the low-energy nonperturbative theory, instanton corrections
to the superpotential {\em always} spontaneously break the gauge group
to its maximal torus; there cannot be solutions for which $\f=0$ at
any point\footnote{Again, we are assuming that even for spacially
dependent fields, it makes sense to consider $\f(u)$ a function of $u$
only.}.  We denote the regions where the quantum corrected field
configurations have unphysical values as the ``hard'' quantum core.
In the case of a classical $(n_m,0)$ monopole solution, there are
usually associated $n_m$ zeros of the scalar field; these roughly
describe the multiple cores of the soliton (however, in some cases the
multi-monopole fields have additional symmetries leading to a higher
number of zeros than expected). Consequently, the quantum-corrected
multi-monopole solutions will have multiple hard cores.

Secondly, in the $SU(2)$ theory there is a region
in the moduli space of vacua where the nonabelian part of
${\ci}m\FF_{AB}$ becomes negative \cite{lgr}; upon crossing into this
region, 
certain BPS states disappear from the semi-classical spectrum
\cite{sw,lgr,bilal}.
Explicitly,  the $\f$-dependent gauge couplings for the $SU(2)$ theory are
defined by the imaginary parts of
\be{HiEtau}
\tau_{AB} = {\partial^2\over \partial\f^A\partial\f^B} \FF(\f)  \ .
\eq
As $\FF$ is a function only of the gauge invariant
quantity $\f=\sqrt{\f^2}$, we have \cite{lgr}
\be{ProjF}
\FF_{AB} (\f) = {\f_A\f_B\over\f^2} \FF\prr
+ (\delta_{AB}-{\f_A\f_B\over\f^2}) {\FF\pr\over\sqrt{\f^2}} \ ,
\eq 
where primes denote derivatives with respect to $\f$.  This leads to a
decomposition onto the $U(1)$ and $SU(2)/U(1)$ field components.  The
$U(1)$ component of the superpotential, ${\ci}m\FF''$, is identified
in the Seiberg-Witten formulation as a positive metric on the moduli
space of vacua (in the $u$ coordinate).  The $SU(2)/U(1)$ fields give
a contribution to the covariantized action
\be{CosetAction}
S_{SU(2)/U(1)} = -\frac1{8\pi} {\ci}m \int d^4x \Bigl({\f_D\over \f}\Bigr)
 {\cal P}_{AB} \Bigl\{ (B_i^A+iE_i^A)(B_i^B+iE_i^B) +
2\nabla_\au \f^A \nabla^\au {\bar\f}^B\Bigr\} \ ,
\eq
where ${\cal P}_{AB}\equiv(\delta_{AB}-{\f_A\f_B/\f^2})$ is the
projection operator onto $SU(2)/U(1)$ (${\cal P}=\Pi$ when $\f^A$ is
radial). The combination ${\tilde\tau}=\f_D/\f$ transforms under
duality transformations in the same manner as $\tau={\del\f_D/\del\f}$
but does not obey any positivity condition; the curve described by the
vacuum expectation values ${\ci}m (a_D/a)=0$ is a real co-dimension
one surface in the quantum moduli space and is topologically a circle
\cite{Ansar,ats}.  Physically, on this boundary the masses of the
fundamental fields become degenerate with monopole/dyon pairs; for the
pure $SU(2)$ gauge theory this curve is closed and separates the
moduli space into two independent sectors that are never mixed under
$SL(2,Z)$ duality transformations. The coefficient of the massive
vector boson kinetic term is negative inside the circle, when
${\ci}m(\f_D/\f)<0$; this signals that the massive vector bosons
disappear from the full nonperturbative theory \cite{lgr}.
 \begin{figure}[htb]
  \begin{center}
 \mbox{\epsfysize=7.0cm \epsfxsize=7.0cm
 \epsfbox[0 0 450 360]{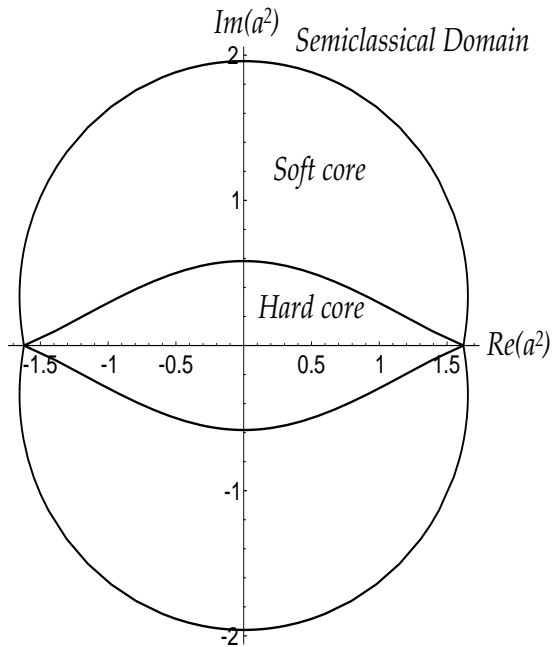}} 
  \end{center}
\vskip .5cm
  \caption{Cores in the $\phi^2$-plane}\label{a2plane}
 \end{figure}

Whenever ${\ci}m \bigl(\f_D/\f\bigr)$ is less than zero, the BPS
equations need not hold because the Hamiltonian (\ref{almostBPS}) is
no longer positive definite.  We denote these regions in the field
solutions as the ``soft'' quantum cores.  The soft core and the hard
core are plotted in the complex $\f^2$-plane in figure \ref{a2plane};
they are of course well known in the $\tau=\FF''$ and
${\tilde\tau}={\f_D/\f}$ plane \cite{ats}.  In this region the soliton
solutions require a better quantum description.  This is not
surprising--we are using a low-energy approximation to the effective
Hamiltonian, and in the core region higher derivative terms that we
are ignoring are certainly not negligible.  However, we know from the
supersymmetry algebra that the BPS mass (\ref{bound}) is preserved
\cite{sw,ow}.  Consequently, the corrections cannot do too much
violence to the solutions.

We have assumed that the gauge-covariantization of the low-energy
effective action $\FF$ is meaningful outside the curve of marginal
stability, and have used this curve to define the soft quantum
core. As the Hamiltonian remains positive precisely up to this curve,
this seems to us a very plausible assumption.  One might argue that
the gauge-covariantized low-energy effective action is meaningful only
when the gauge bosons are the lightest states in the theory. This
would define a ``Wilsonian'' quantum core by the condition $\vert
\f_D/\f\vert \le 1$ or $\vert \f_D/\f\pm1\vert \le 1$. The Wilsonian
core lies somewhat outside the soft core. Though our subsequent
discussion of the boundary conditions on the surface of the quantum
core would be modified if we assumed the solutions were valid only up
to the Wilsonian core, the rest of our results would remain
unchanged.\footnote{We are happy to thank Phil Argyres for discussions
on this point.}

\section{Solutions}

In this section we discuss several types of solutions to the BPS
equations (\ref{BPS}).  We describe how the boundary conditions of the
fields on the soft quantum core lead to two different quantum analogs
of the classical 't Hooft-Polyakov monopole.

The condition ${\ci}m(\f_D/\f)=0$ at the boundary of the
quantum core imposes constraints on the fields $X$ and $X_D$.
The solutions to (\ref{scalarbox}) and (\ref{dualbox}) (for
$n_m=\pm1$) are degenerate
in the quantum corrected theory because of the identity
\be{rPhase}
n_e X(r)+n_m X_D(r)=0  \ ;
\eq
this means that the phase of $Z(r)$ is constant.
Consequently, at the boundary of the soft core, denoted by the radius $r_c$,
\be{XnZpoles}
0 = {\ci}m \left( {\f_D(r_c)\over \f(r_c)}\right)
&=&\frac1{n_m\f(r_c)\bar\f(r_c)}
{\car}e\left(in_m\f(r_c)\bar\f_D(r_c)\right)\nonumber\\
&=&\frac1{n_m\f(r_c)\bar\f(r_c)}
{\car}e\left(i\f(r_c)\bar Z(r_c)\right)= 
{1\over n_m \f{\bar\f}} X(r_c) \vert Z(r_c)\vert \ ,
\eq
where we use $X(r)={\car}e\left(i\bar Z\f(r)/\vert Z\vert\right)$.
Thus either the field $X(r_c)$ or the $r$-dependent central charge
$Z(r_c)$ vanishes at the boundary $r_c$ of the soft core, and there are
two different types of dyonic solutions which we call $X$-poles and
$Z$-poles. We shall see below that the distinction between $X$ and
$Z$-poles has a global significance that is not tied to the boundary
conditions at the core.

When $Z(r_c)=0$, we know that $u(r_c)=\pm1$. Thus $Z$-pole solitons
are $r$ dependent solutions described by curves in the $u$-plane
running from some vacuum value of $u$ and to $u(r_c)=\pm1$, depending
on the quantum numbers: $(n_m,n_e)=(1,0)$ $Z$-poles approach $u=1$ at
the core, and $(1,\pm1)$ $Z$-poles approach $u=-1$. The $X$-poles have
solutions where $X(r)=0$ before reaching the $u=\pm 1$ points. Thus,
at the critical radius, $X(r_c)=0$, and from equation
(\ref{radialBPS}) we find that $({d\over dr} \ln{L}) \vert_{r_c} =0$.
Within the spherical ansatz, equation (\ref{berad}) then implies that
the nonabelian part of the $B$-field ({\it i.e.}, the $\Pi_{AB}$
term) vanishes at $r_c$.  Note that $Z$-poles have no contribution on
the surface of the quantum core, and hence no net internal mass as
well. In contrast, since $Z(r_c)$ is nonvanishing at the core
boundary for $X$-poles, their core has a
positive mass.

We now discuss the solutions by analyzing the constraint on the phase
of $Z$ in equation (\ref{rez}).  In the $n_e=0$ sector, the $r$
dependent central charge is simply $Z(r)=\f_D(r)$, and the lines of
constant $Z(r)$ phase are straight lines from the origin $\f_D=0$
({\it i.e.}, $u=1$) in the $\f_D$ plane. The phase associated with
each $Z(r)$ line is set by the vacuum expectation value
$a_D(u_\infty)$ of the particular soliton solution.  These straight
lines therefore represent soliton solutions as a function of position
space through their $u$ dependence (i.e. from $u(r_c)=1$ to
$u(r=\infty)$), subjected to the boundary conditions of equation
(\ref{XnZpoles}).
\begin{figure}[htb]
  \begin{center}
\mbox{\epsfxsize=7.2cm \epsfysize=10.0cm \epsfbox{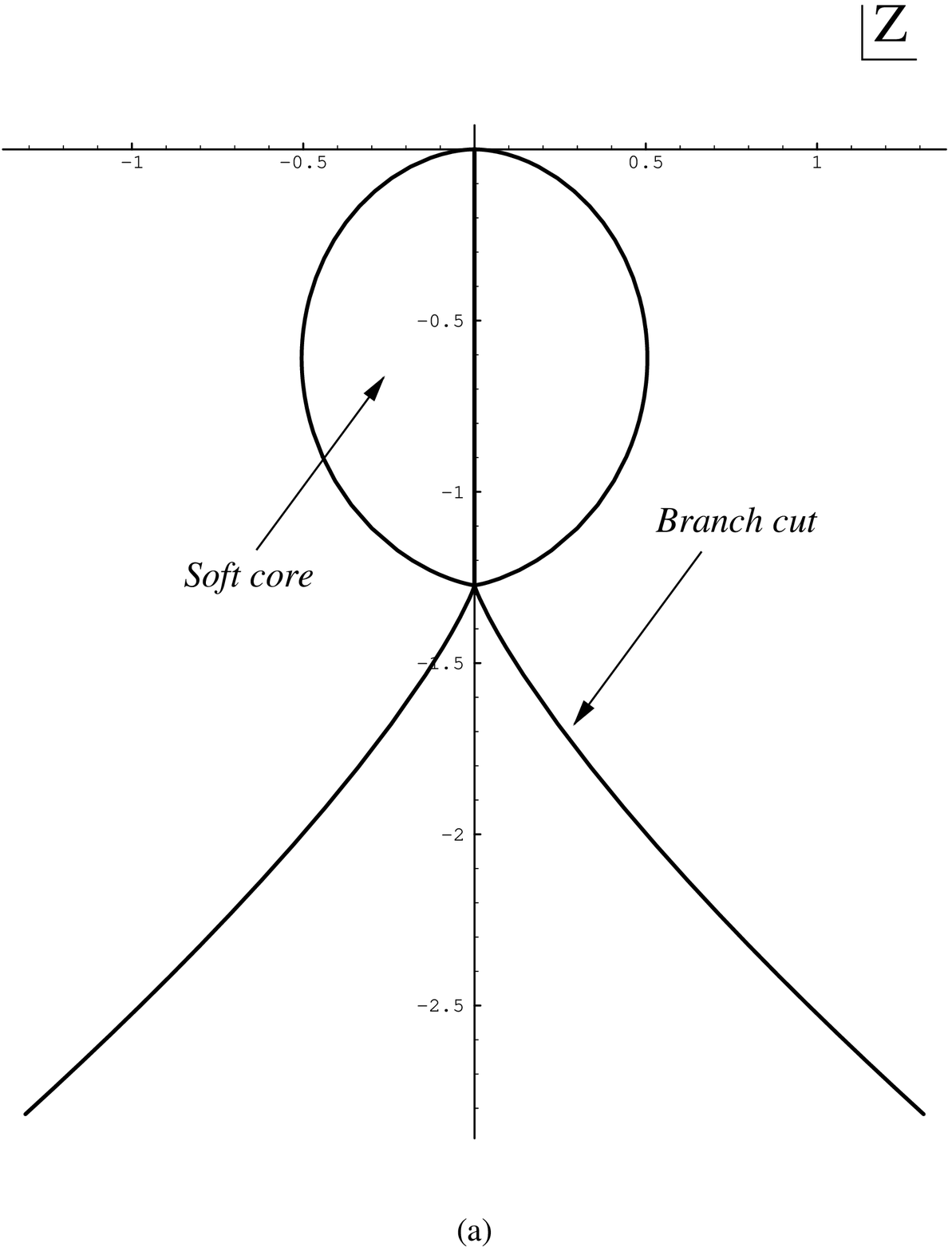}}\hskip 1.0cm
\mbox{\epsfxsize=7.2cm \epsfysize=10.0cm \epsfbox{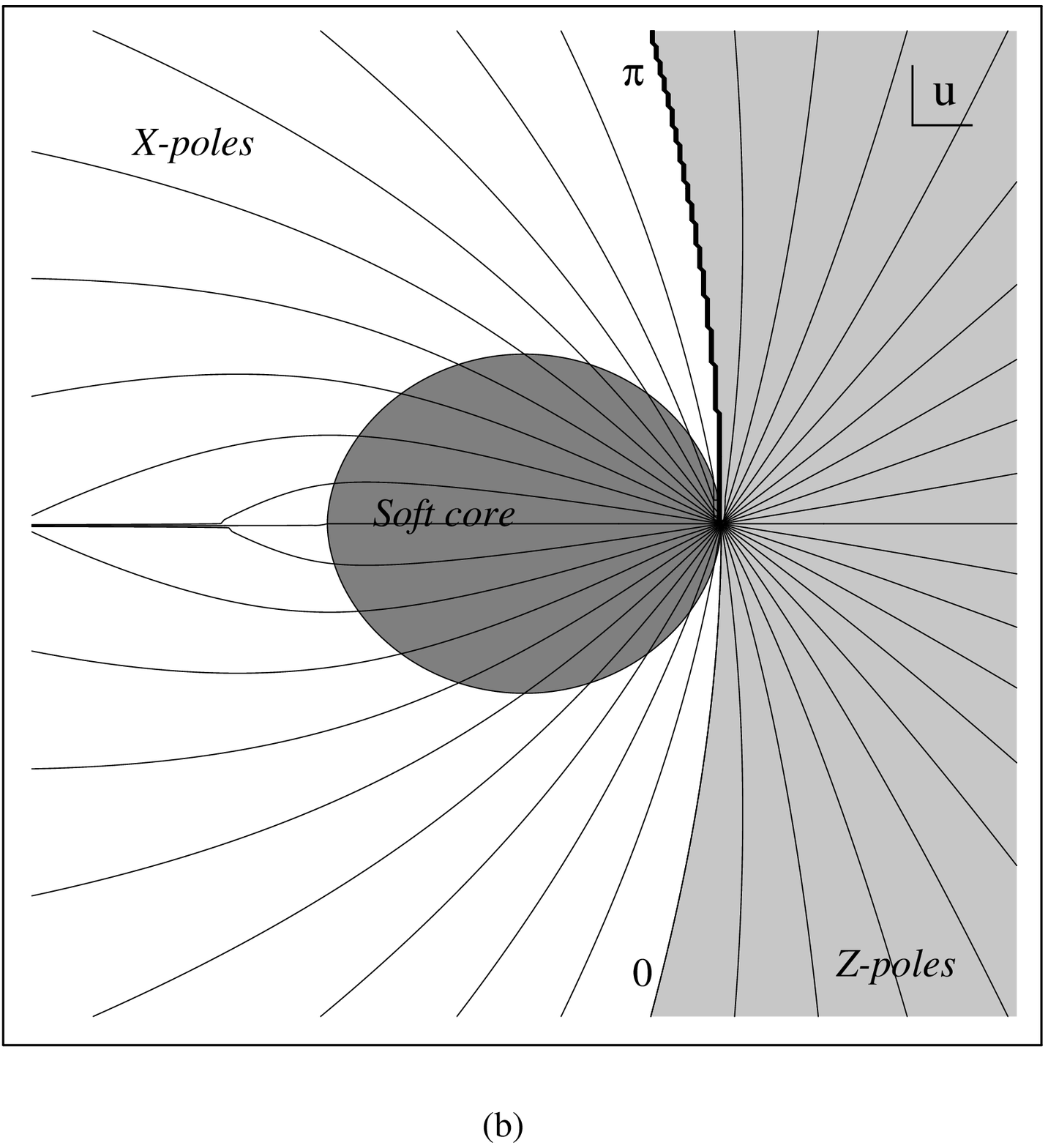}}
  \end{center}
\vskip -.5cm
  \caption{(a) The core and branch cut in the $Z=\f_D$-plane.
(b) Lines of constant $Z$-phase in the $u$-plane are shown for
  $\pi/20$ increments.}\label{manylines}
 \end{figure} 

The $\f_D$-plane is shown in figure \ref{manylines}(a): The closed curve
is the boundary of the soft core, and the two curves running out from
the bottom of the soft core are the map to the $\f_D$-plane of a cut
along the negative $u$-axis starting from $u=-1$.  Beneath these
curves are values of $\f_D$ that are forbidden because of the
discontinuity in $\f_D$ as one crosses the cut.

Lines of constant $Z(r)$-phase in the $\f_D$ plane have two types of
behaviours characterized by their slopes: When the vacuum expectation
value $a_D(u)$ lies in the upper half of the complex plane, $Z(r)$ can
hit the soft quantum core region only at $u(r_c)=1$, where
$Z(r_c)=\f_D =0$, corresponding to a $Z$-pole.  When $a_D(u)$ is in
the lower half-plane, the line intersects the soft quantum core
boundary at some critical radius $r_c$ where $X(r_c)=0$ (and
$Z(r_c)\neq 0$), corresponding to a $X$-pole. In the latter case the
solution breaks down within the soft quantum core, although one may
formally continue it all the way to the $u({\tilde r}_c)=1$ point
where we may define the second critical radius ${\tilde r}_c$.

In the $\f_D$-plane, the curves representing the branch cut tend to
horizontal lines, albeit only logarithmically.\footnote{The limit
of $\ci{m}(\f_D)/\car{e}(\f_D) = 1/\ln(-u)$ for $u$ real, large, and
negative.}  Thus an $X$-pole solution has a maximum allowed value of
$\vert Z(\infty)\vert$; continuing across the cut gives rise to
solutions related by a monodromy transformation $M_\infty$ (see
\ref{mono}), and we shall find these solutions when we consider the
$(n_m,n_e)=(1,2)$ sector. This gives an a priori independent
definition of $X$-poles and $Z$-poles that is not based on boundary
conditions at the core of the monopole: analytically continuing to
larger mass along lines of constant $Z$-phase, $X$-poles eventually
reach another charge sector.  Note that for $X$-poles and $Z$-poles,
the phase of $Z(r)$ (and hence of $\f_D(r)$) is known from the
asymptotic value; however, the explicit solution to $\f_D(r)$ involves
solving a nonlinear integro-differential equation for $u(r)$
(\ref{udependence}).  (We present an alternative formulation for the
solution to $u$ through a nonlinear differential equation in Appendix
A.)

In the $u$-plane, the lines of constant $Z$ phase are drawn in figure
\ref{manylines}(b) together with the soft quantum core (shown shaded
darkly).  There are two types of lines corresponding to the different
solutions described above.  All of the lines in the $u$-plane extend
outwards from the $u=1$ point.  The lines in the lightly shaded region
represent $Z$-poles; clearly these arise only for certain values of
$u(\infty)$.  The lines in the unshaded region represent
$X$-poles. These eventually hit the negative $u$-axis, as described in
the previous paragraph.  Continuing $u$ through the negative real axis
brings the solutions through the branch cut and implies a monodromy
transformation; the continued solutions beyond the cut represent
$X$-poles in a different charge sector.

As mentioned in section 3, the quantum corrected solutions have a
remarkable feature: the $\Theta$-angle is spatially dependent! For
$Z$-poles, it bleaches out as we approach the core, whereas for
$X$-poles, it can either bleach or intensify depending on the value of
$u(\infty)$. This may be seen simply by noting that the solutions as a
function of ${\vec r}$ come out of the $u=1$ point to some vacuum
value $u(\infty)$.  The coupling constant $\tau(u)$ must change along
the solution for $u({\vec r})$ as a function of ${\vec r}$.  Recall
that the $\Theta$-angle is determined by ${\car}e \left(\tau\right)$.
This phenomenon has no analog classically, as the classical
$\Theta$-angle is a fixed parameter. However, it is known to occur
quantum-mechanically in the presence of light charged
fermions.\footnote{We thank A.\ Goldhaber for discussions on this
point.} A possible interpretation of the phenomenon is that near the
quantum core, the ``local'' mass of the $(n_m,n_e)=(1,0)$ dyons is
going to zero, and hence the dyons themselves might act as the light
fermions that bleach the $\Theta$-angle.
\begin{figure}[htb]
\begin{center}
 \mbox{\epsfxsize=9.0cm \epsfysize=9.6cm
  \epsfbox[0 0 471 500]{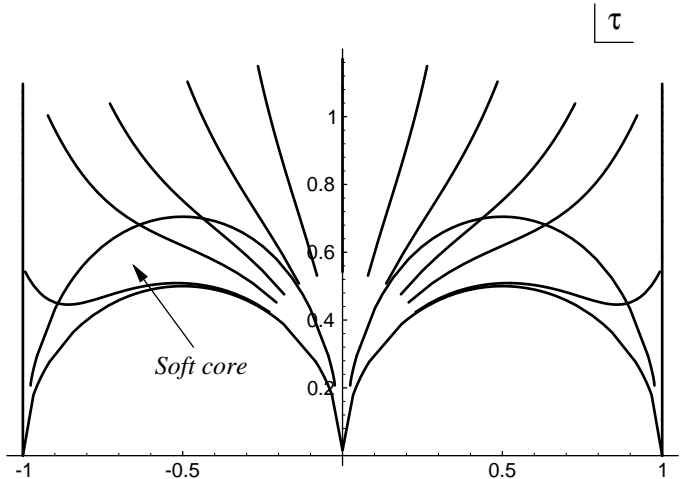}} 
  \end{center}
\vskip -4.0cm
  \caption{Lines of constant $Z(r)$-phase in
the $\t$-plane in the $n_e=0$ sector.}\label{adplane}
 \end{figure}

We illustrate the bleaching phenomenon by displaying several monopole
solutions in the $\tau$-plane in figure \ref{adplane}.  The origin
corresponds to $u=1$ and the points $\pm 1$ correspond to $u=-1$.  The
two lower half circles in the $\t$-plane are the map of the real line
segment $u\in [-1,1]$ (the boundary of the hard core) and the two
above them come from the map of the ${\cal I}m\Bigl(a_D/a \Bigr) =0$
curve (the boundary of the soft core) into the $\tau$-plane
\cite{ats}.  Several soliton solutions ({\it i.e.}, lines of constant
$Z(r)$ phase) in the $\tau$-plane are displayed.  One sees clearly
that the $\Theta$-angle changes along the lines.  The $Z$-pole and
$X$-pole are distinguished in this plot as usual by whether or not
their solutions cross through the curve of marginal stability ${\cal
I}m\Bigl(a_D/a \Bigr) =0$.
\begin{figure}[htb]
  \begin{center}
 \mbox{\epsfysize=10.8cm \epsfxsize=7.2cm \epsfbox{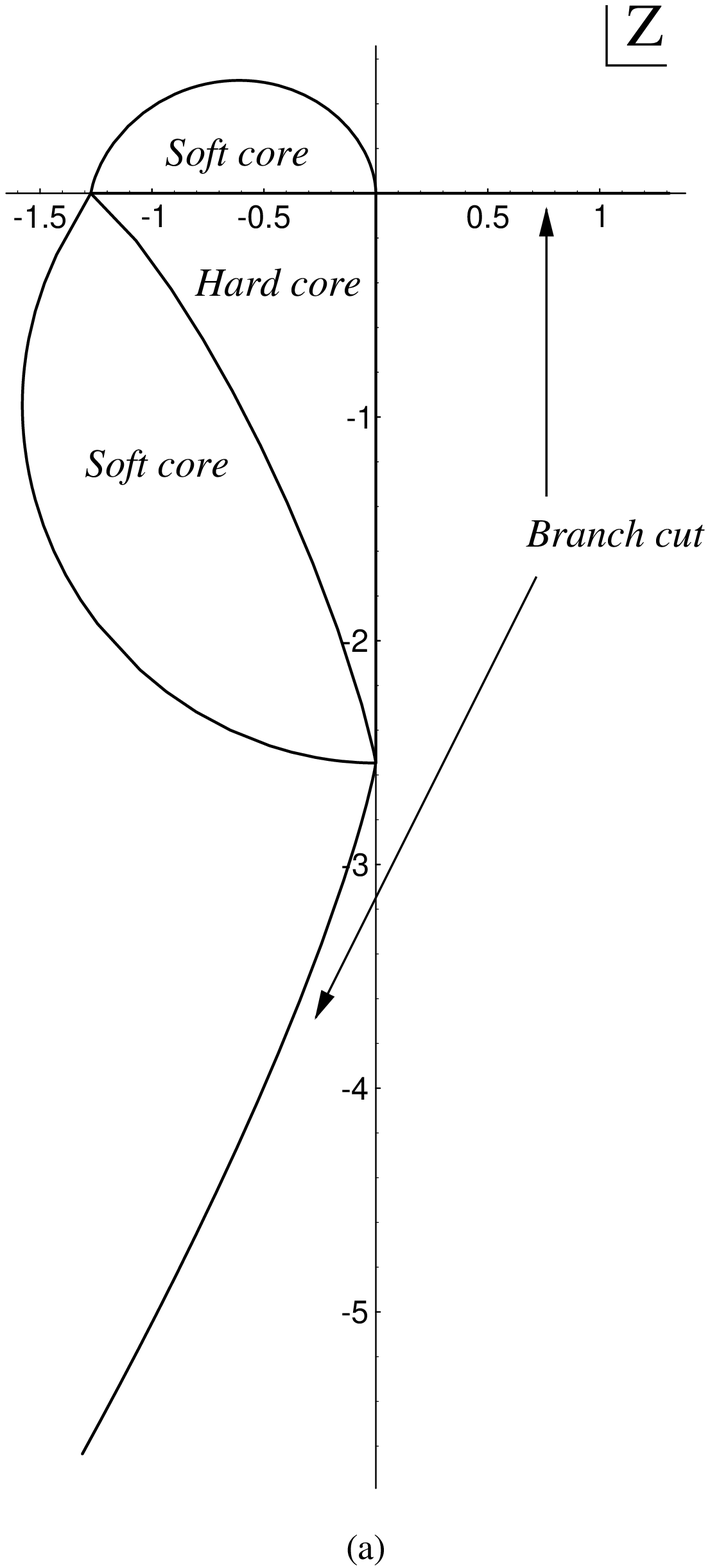}}\hskip 1.0cm
 \mbox{\epsfysize=10.0cm \epsfxsize=7.2cm \epsfbox{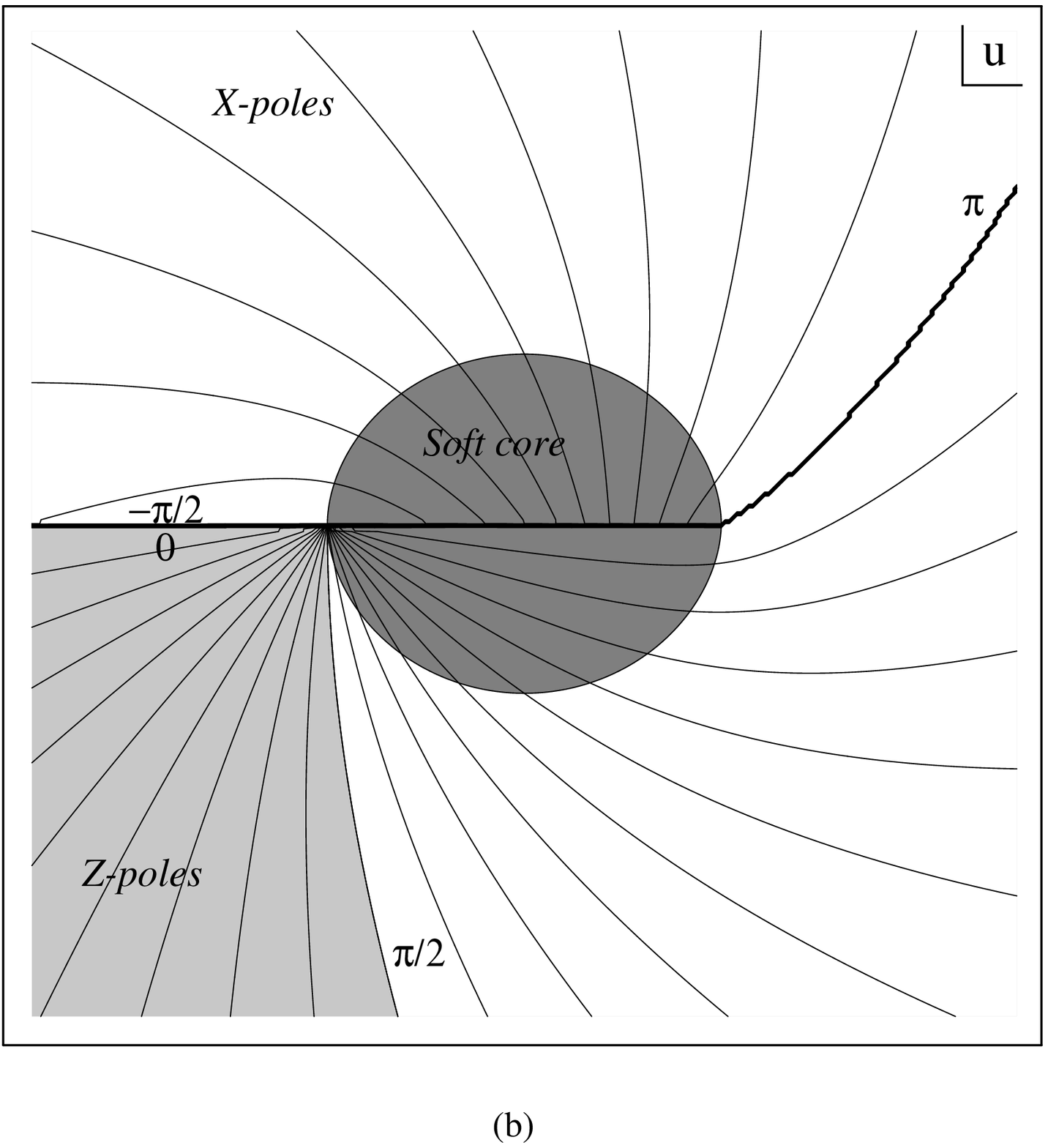}}
  \end{center}
  \caption{Sample solutions in $n_e=-1$ sector. Lines of constant
  $Z$-phase in the $u$-plane are shown for 
  $\pi/20$ increments.}\label{nem1}
 \end{figure} 

\begin{figure}[htb]
  \begin{center}
 \mbox{\epsfysize=9.6cm \epsfxsize=9.0cm \epsfbox[0 0 471 500]{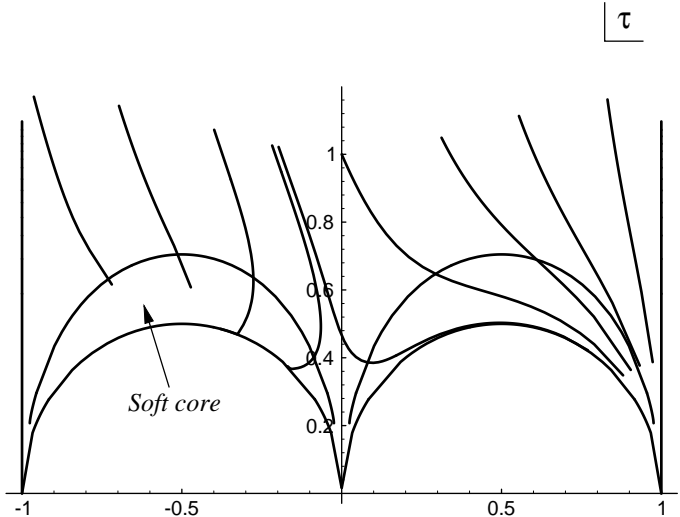}}
  \end{center}
\vskip -4.0cm
  \caption{Lines of constant $Z(r)$-phase in the $\t$-plane 
  in the $n_e=-1$ sector.}\label{taum1}
 \end{figure}

\begin{figure}[htb]
  \begin{center}
 \mbox{\epsfysize=10.6cm \epsfxsize=7.2cm \epsfbox{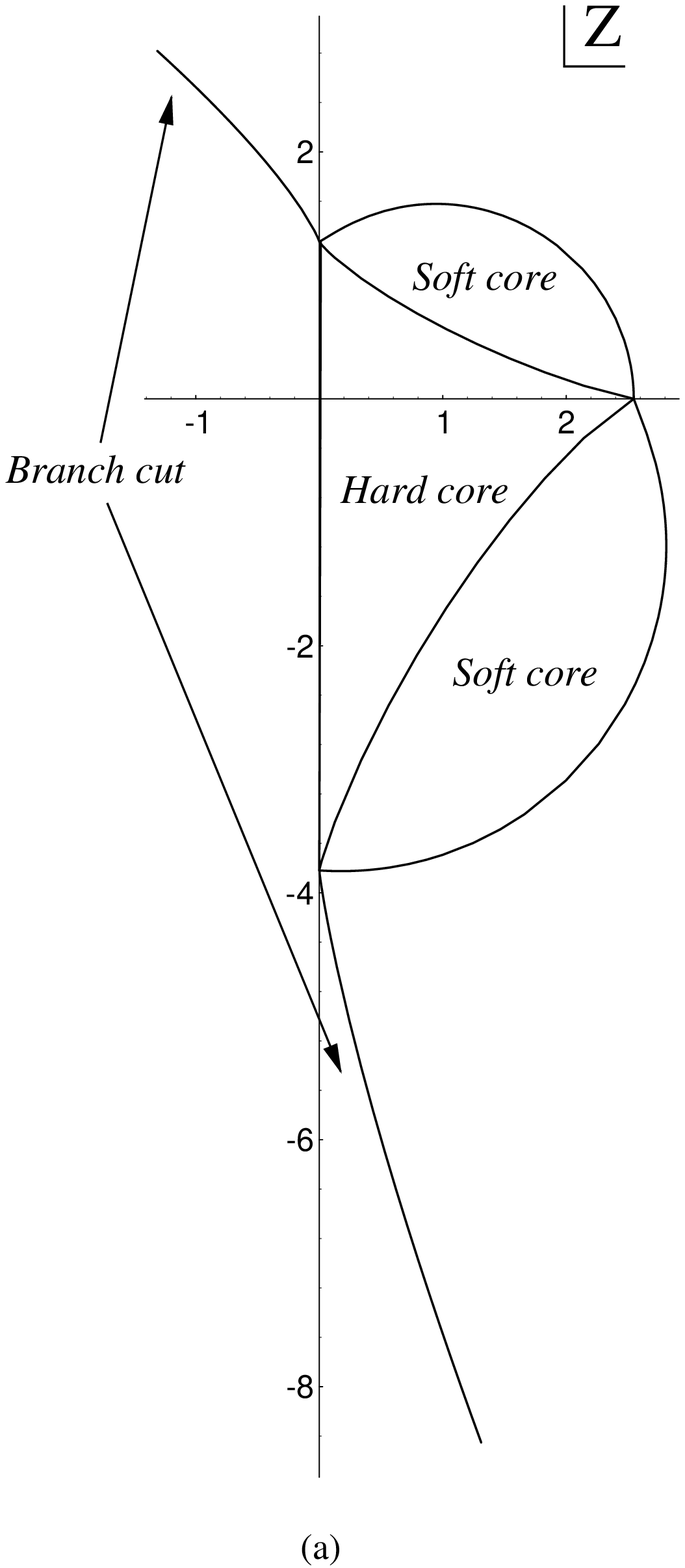}}\hskip 1.0cm
 \mbox{\epsfysize=10.0cm \epsfxsize=7.2cm \epsfbox{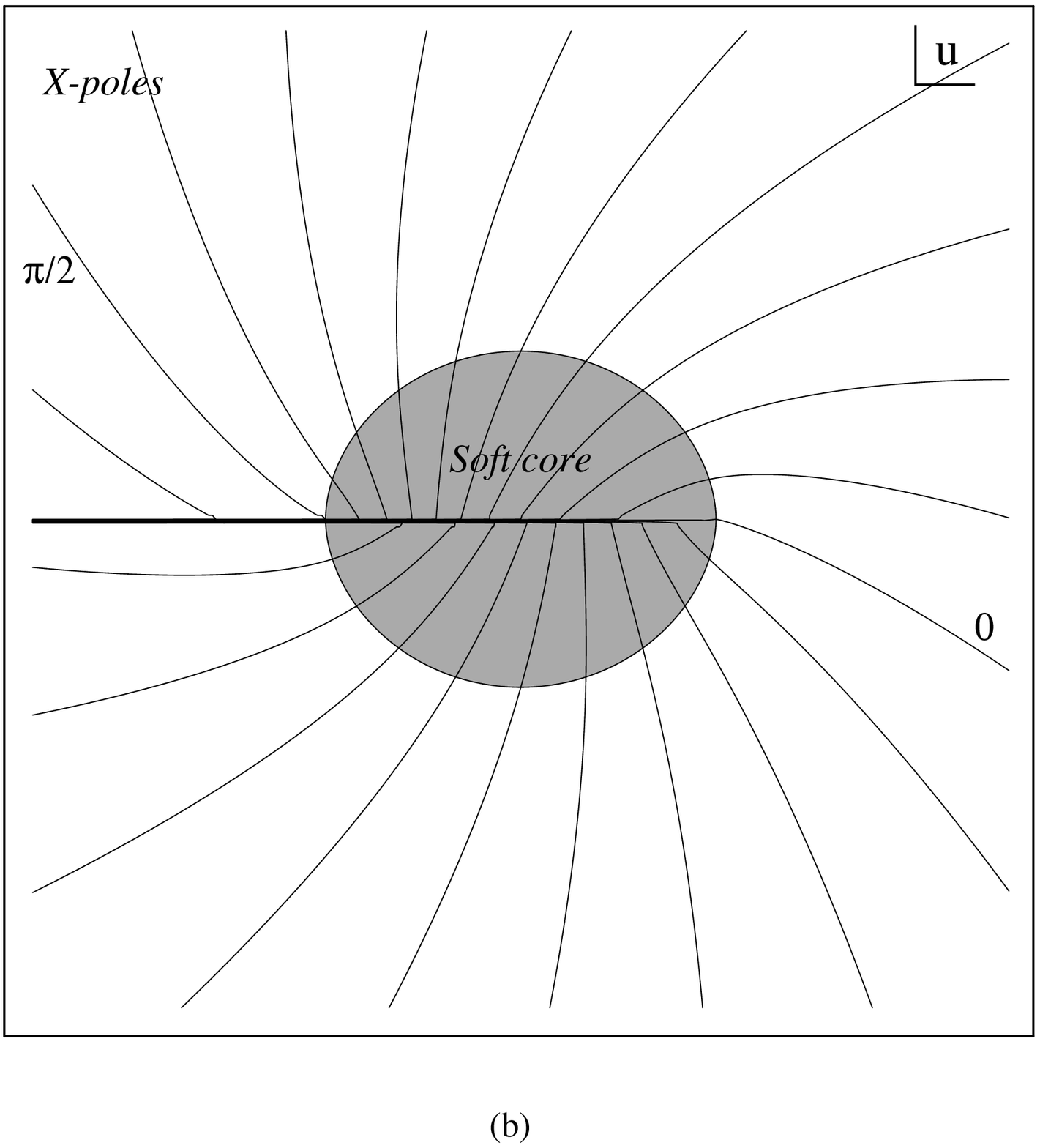}}
   \end{center}
  \caption{Sample solutions in $n_e=2$ sector.  Lines of constant
  $Z$-phase in the $u$-plane are shown for 
  $\pi/20$ increments.}\label{ne2}
 \end{figure}

In the $n_e=-1$ sector the structure of the solutions is the same and
is again found by following radial lines in the $Z$-plane; however, in
this case the $Z(r_c)=0$ condition appears at the $u(r_c)=-1$ point.
Sample solutions in the $Z$, $u$, and $\t$ planes are illustrated in 
figures \ref{nem1} and \ref{taum1}.  Further examples for $n_e=2$ are 
given in figure \ref{ne2}; note that in this charge sector, all
solutions are $X$-poles.

The remaining issue to be discussed is the effect of monodromies in
$a(u)$ and $a_D(u)$ on the quantum corrected solutions.  Because of
the breakdown of the BPS bound within the quantum core our analysis is
restricted to the semi-classical monodromy: encircling one of the two
points $u=\pm 1$ involves passing through the quantum core, about
which we are ignorant.  The semi-classical monodromy on the solutions
is obtained from $M_\infty$ in (\ref{mono}).

We may consider a solution for some asymptotic $u(\infty)$, {\it
e.g.}, in the $Z$-pole region.  Continuing $u(\infty)$ clockwise, the
solution becomes an $X$-pole, and eventually $u(\infty)$ reaches the
negative $\car{e}(u)$-axis. We then leave the $n_e=0$ sector and cross
over to the $n_e=2$ sector; indeed, we see that the phase $-9\pi /20$
contour in figure \ref{manylines}(b) continues as the phase $11\pi
/20$ contour in figure \ref{ne2}(b) (the $\pi$-phase difference is due
to the change of sign of $n_m$ in (\ref{monod})).

\section{Spatial dependence}

The general solution in (\ref{generalL}) depends on a parameter $\d$
which is presumably fixed by the microscopic theory in the quantum
core region.  Classically, regularity of the fields at the origin
imposes $\d=0$.  We comment here on some of the effects of the $\d$
parameter on the soliton solutions within the semi-classical region.
For simplicity, we only consider the $n_e=0$ sector.

Asymptotically, at large $r$ the scalar field $X$ behaves as
\be{xas}
\sqrt{2} X(r)= \k - {1\over r} + 2\k e^{-2\k(r+\d)}+\ldots \ .
\eq
The first two terms describe just the vacuum expectation value of the
scalar field and the topological magnetic charge.  The exponential
correction reflects the presence of the massive vector bosons in the
theory; the decay constant $\k=\sqrt2{\car}e\left(\exp(i\a)a\right)$
is proportional to their mass ($\sqrt2\vert a \vert$).  The $\d$
parameter does {\em not} affect the rate of fall-off; rather, it
suppresses the exponential correction by a factor $e^{-2\k\d}$.
Physically, this suggests that $\d$ reflects the absence of the
massive vector bosons in the quantum core, and a resulting suppression
of their contribution to the monopole. Indeed, taking $\d\to\infty$
leads to a Coulomb form for the magnetic and electric fields in
(\ref{berad}).

The critical radius $r_c$ for $X$-poles is determined by the
condition $X(r_c)=0$:
\be{xcon}
\sqrt2X(r_c)=\k \coth{\k(r_c+\d)}-{1\over r_c} = 0 \ .
\eq
This implies that for $X$-poles, $r_c$ is {\em always} less than the
classical core radius $R\equiv1/\k$ for any $\d$, with $r_c\to 0$ when
$\d\to0$ and $r_c\to R$ when $\d\to\infty$. Note that as $X(\infty)$
tends to the curve of marginal stability, the classical radius
$R=1/(\sqrt2X(\infty))$ as well as $r_c$ tend to $\infty$.

For $Z$-poles in the $n_e=0$ sector, $Z(r_c)=0$ implies
\be{zcon}
\sqrt{2} X(r_c)=
 \k\coth{\k(r_c+\d)} -{1\over r_c} =
 \sqrt{2} {\car}e\Bigl[ e^{i\a} \f(u=1) \Bigr] =
\frac{4\sqrt2}\pi\cos{\a}\  ,
\eq
where $\cos{\a}=\sin{\left(arg(Z)\right)}$, and is always positive, as
$n_e=0$ $Z$-poles occur when $a_D$ is in the upper half-plane (see
figure \ref{manylines}a). The value $\f(u=1)=4/\pi$ is an instanton
induced constant that is proportional to $\L$ (we have set $\L=1$
throughout the paper).  In the limit of vanishing $\L$, $r_c\to0$,
and the monopole becomes classical.
\begin{figure}[htb]
  \begin{center}
 \mbox{\epsfxsize=9.0cm \epsfysize=9.0cm \epsfbox[0 0 500 500]{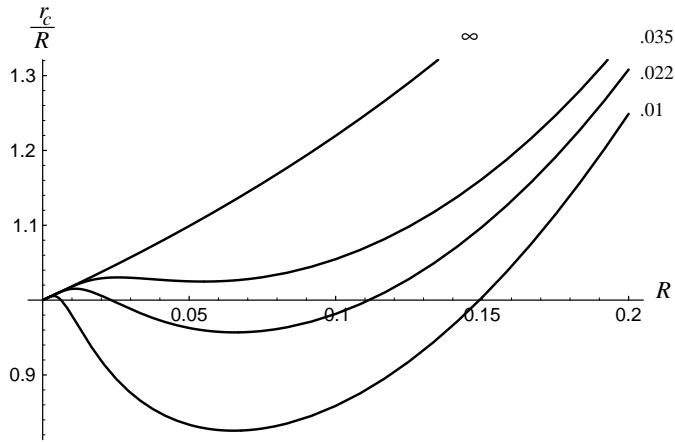}}
  \end{center}
\vskip -3.5cm
  \caption{The ratio $r_c/R$ as a function of $R$ for
 $\d=\{.01,.022,.035,\infty\}$ and $\a=0$.}\label{rcfig}
 \end{figure}

Whereas for $X$-poles $r_c\le R$, for $Z$-poles, the quantum core
radius $r_c$ may be less than or greater than the classical core
radius $R$, depending on the values of $R$, $\d$ and $\a$. In figure
(\ref{rcfig}), the $r_c/R$ is plotted against $R$ for various values
of $\d$, with $\a=0$.  For $Z$-poles, as $u(\infty)\to1$, the
classical radius $R=1/(\sqrt2X(\infty))$ remains finite whereas the
quantum radius $r_c\to\infty$.  For $Z$-poles with $n_e=\pm1$, the
situation is very similar, with $\cos\a\to\pm\sin\a$ above, and the
phase of $Z$ is restricted to the quadrant that ensures
$\car{e}(\exp(i\a)\f(r_c))=\lim_{\vare\to0}\car{e}(\exp(i\a)
\f(u=-1\pm\vare))$ is positive (see figure \ref{nem1}(a) for $n_e=-1$;
the analogous figure for $n_e=1$ is simply the reflection about the
$\ci{m}(Z)$-axis).

\section{Next Order Corrections}

In the low-energy theory described by the superpotential $\FF(\phi)$
we have shown how BPS saturated states arise and generalize the
classical soliton solutions.  However, the superpotential $\FF(\f)$ is
just the first term in a momentum expansion of the low-energy
effective action.  There are higher derivative corrections; the first
term is a real potential $\HH$ of dimension zero integrated over the
full $N=2$ superspace measure (\ref{Hdef}).  In this section we find
how these higher order corrections affect the quantum solitons.
Contributions to $\HH$ of two different kinds have been discussed
\cite{h,HH1,HH2,matone}.
 
In \cite{HH1} it was found that the one-loop correction to the
K\"ahler potential associated to the {\em nonabelian} low-energy
theory $L_\FF$ cannot be written in the form ${\ci}m {\bar\f} \FF_\f$
unless $\f$ and $\bar\f$ commute. This is in direct conflict with the
constraints imposed by special geometry.  However, the next order
correction $\HH(W,{\bar W})$ discussed in \cite{HH1} contributes to
the nonabelian part of the low-energy theory and repairs the apparent
inconsistency.  Here, we focus on this particular contribution, as it
gives rise to corrections to the K\"ahler potential (albeit,
corrections that vanish in the abelian limit). 

The correction $\HH$ we take is a function of the parameter  
\be{tpar} 
t={\f\cdot{\bar\f}\over \sqrt{\f^2{\bar\f}^2}} \ .
\eq 
The parameter $t=1$ when the fields are restricted to be 
abelian.  All scalar derivative terms of the $\HH$ contribution 
when written as a function of $t$ are given in appendix B.

In addition to the nonabelian correction above, an explicit form for
the {\it abelian} correction has been proposed in \cite{matone}:
\be{hmat} 
\HH(\f,{\bar\f})=4\pi^2 \vert 
\frac{\del\f}{\del u}\vert^4 \vert u^2-1\vert 
({\ci}m \tau)^2 \ .
\eq 
This term is expressed as a function of the dimensionless group
invariant $x=\f^2/\L^2$, in contrast to the parameter $t$ in
(\ref{tpar}).  As a result, the scalar derivatives of $\HH$ in this
form will be quite different.  For example, whereas the $\HH_A$ term
vanishes for the nonabelian correction when expressed as a function
of $t$, it contributes to the full $U(1)$ term.  Higher derivatives
are also different, and in some sense the two corrections are
orthogonal to one another. We have not investigated the contribution
of this term to the dyon mass.

The contribution from the $\HH$ function in (\ref{Hdef}) to the
Lagrangian when expanded out into $N=0$ components\footnote{There is
an ambiguity in defining the $N=2$ superspace measure which consists
of total derivatives of functions of the fields.  In deriving the
energy we have used the definition of the $\HH$-function as listed in
appendix B and have not integrated by parts, which on an individual
$N=0$ term may in introduce possible singularities at the origin.  We
have taken the form in (\ref{Hdef}) as the definition of the
$\HH$-function contribution.} contains many higher derivative terms,
and we refer to appendix B for the explicit calculation.  Previously
we found the canonical Hamiltonian associated to the superpotential
$\FF$; because of the higher derivative terms, it is awkward to define
the Hamiltonian canonically.  Rather, we evaluate the time-like
components of the energy-momentum tensor derived from the full
Lagrangian.  A general term $S_i$ in the action with $n$ indices may
be written as
\be{general}
S_i = \int d^4x \sqrt{-g}g^{\au_{1}\bu_{1}}\ldots g^{\au_{n}\bu_{n}}
T_{\au_{1}\bu_{1}}\ldots\mbox{}_{\au_{n}\bu_{n}}
\eq
where we take a mostly positive Minkowski metric.  By varying with
respect to the metric and looking at the 00-components we find the
energy contribution from this term as
\be{saas}
E=\half T_{i_{1}i_{1}}\ldots\mbox{}_{i_{n}i_{n}}+
\half\left(T_{00\;i_{2}i_{2}}
\ldots\mbox{}_{i_{n}i_{n}}+{\rm perms}. \right)
-\frac{3}{2}\left(T_{00\;00\;i_{3}i_{3}}\ldots\mbox{}_{i_{n}i_{n}}
+{\rm perms}.\right) +\ldots
\eq
Terms with $\e$ tensors must be treated differently since they
transform as densities.  All of these contributions vanish when $\f$
is real; the electric field is vanishing in this case and no nonzero
contractions of the field strength and its dual are possible.

For simplicity we only consider radially symmetric soliton backgrounds
and we investigate their contribution to the energy.  The gauge and
scalar fields in the radial ansatz are given in (\ref{radans}), and we
find the contribution to the energy of the soliton by substituting
their form into the expansion of the energy as derived from $\HH$.  In
the radial ansatz the parameter $t\equiv1$ throughout all
space, and the net contribution is expressed in terms of the two
constants $\HH'$ and $\HH''$ presumably determined by the nonperturbative
structure of the low-energy theory.  Details of the full calculation
from all the terms in the $N=2$ superspace expansion are given in
appendix B.

The contribution from $\HH(W,{\bar W})$ to the energy of the
radially symmetric field configuration is given below.  For
ease of reading we group the terms into four contributions:
\be{FullEnergy}
E_1 &=& 2\pi (2\HH''+\HH') \int dr~ \frac{1}{\bigl|\f^4\bigr|}
  \left(\frac{L^2}{r}\right)^2 \Bigl\{ \left( \bigl|\f\bigr|^2 +\half
  b^2- \half\left(\frac{L_r}{L}\right)^2\right)^2 - b^4
 \Bigr\} \ ,
 \nonumber\\
E_2 &=& \pi \HH' \int dr~ {1\over \vert\f\vert^2} \vert L_r
{\f_r\over\f}
 + {L(1-L^2)\over r^2} \vert^2
 - {L^2 b^2\over \vert\f\vert^2} \vert {b_r\over b}-{\f_r\over\f}
 \vert^2 \ ,
\nonumber\\
E_3 &=& 4\pi\HH\pr\int dr~ \left(\frac{Lr}{2}-
       \frac{gr^2}{4\f\bar{\f}}\right) \left\{
2\frac{L(1-L^2)}{r^3} +\frac{L_r}{r}\left(\frac{\f_r}{\f} +
c.c.\right)\right\} \ , \\
E_4 &=& 4\pi\HH\pr\int dr~ \frac{r^2}{\bigl|\f\bigr|^2}\left\{
\left(\frac{g}{2}-\frac{L\bigl|\f
\bigr|^2}{r}\right)^2-\left(\frac{b^2 L}{r}\right)^2\right\}
\ .\nonumber
\eq 
In equation (\ref{FullEnergy}) the function 
$g$ is defined by
\be{gis}
 g \equiv \frac{L_{rr}}{r} + \frac{L(1-L^2)}{r^3}-\frac{Lb^2}{r}  \ ,
\eq
and arises naturally in the covariant derivative of the field strength
in the radial ansatz. 

We find by inserting the quantum dyon solution for the fields $L(r)$,
$\f(r)$, and $b(r)$ in (\ref{generalL}) and (\ref{radialBPS}) into the
energy expression above that both the $\HH'$ and $\HH''$ terms are
individually vanishing.  Remarkably, the {\em classical} monopole and
dyon appears generically in the low-energy effective $N=2$ theory --
including the next order corrections. The BPS mass formula does not
receive further corrections from these terms. Although it seems
astonishing that corrections to the gauge-field kinetic terms
\cite{HH1} do not change soliton masses, this is an expected
consequence of supersymmetry.

\section{Discussion}

We believe that our attempt to understand quantum corrections to
solitons by studying minima of the quantum corrected low-energy
effective Hamiltonian opens up a new area for investigation. We
conclude with a list of open problems.

We would like to understand what happens inside the quantum core and
to determine the unknown parameter $\d$. For $Z$-poles, there seems to
be a promising approach: near $u=1$, there is a dual description in
terms of a fundamental monopole hypermultiplet weakly coupled to the
dual $U(1)$ gauge-supermultiplet $W_D$ (whose leading scalar component
is $\f_D$). Unfortunately, this dual description is infrared free, so
when the quantum core is small, the theory cannot be considered weakly
coupled. It remains to be seen if there is a range of parameters where
a useful description can be found.

It is also interesting to investigate multi-monopole configurations.
Although one cannot use the radial ansatz, we believe that much of our
analysis would survive, and in particular, that the phase of the local
central charge $Z(\vec r)$ would be constant everywhere in space. In
classical multi-pole configurations, there are multiple zeros of $\f$;
correspondingly, we expect that the quantum-corrected multi-poles
should have multiple quantum cores.

The low-energy dynamics of a charge $k$ monopole in a classical
$SU(2)$ Yang-Mills-Higgs theory broken to $U(1)$ is modelled by
geodesic motion on a smooth $4k$-dimensional hyperk\"ahler manifold
\cite{ahbook}. Presumably, quantum moduli spaces also inherit the
hyperk\"ahler property from the BPS equations, and an $SO(3)$ group of
isometries from spatial rotations of the monopoles. For well separated
monopoles (where the vacuum expectation values are such that the
quantum cores are sufficiently small), we expect the metric to be
smooth and very similar to the classical case.  In this limit the
exponential corrections to the classical monopole and moduli spaces are
negligible, and the hyperk\"ahler metric models the motion of point-like
monopoles.  However, to understand the quantum moduli spaces fully, we
need to find the effects of the quantum cores; in particular, do they
spoil the completeness of the moduli spaces or change the topology
near the origin?

In particular, for the classical charge two configuration, two
incoming monopoles may be converted into outgoing electrically charged
dyons \cite{ah,ahbook}. In the quantum theory, monodromies can change
the charges of monopoles. It would be interesting to see if there is a
relation between these facts: How do quantum effects modify the charge
exchange process, and are monodromies involved?

Beyond answering questions about the pure $SU(2)$ theory, it is
interesting to consider a variety of extensions: One can add
fundamental hypermuliplets, or go to higher rank gauge groups. Indeed,
the simplest quantum moduli space to investigate describes distinct
fundamental monopoles in higher-rank gauge groups.  In this case, the
the moduli space in the classical theory is the higher dimensional
analog of the Taub-NUT metric \cite{arbgps,mm,gc} and has neither
exponential corrections nor charge exchange scattering. This
suppression is caused by the existence of a maximal number of
commuting triholomorphic isometries on the moduli space, a feature
that should persist in the quantum theory.

In the low-energy $N=2$ effective action for higher rank gauge groups,
the gauge symmetry is always broken down to its maximal torus.
Recently, it has been shown that in the classical theory with
nonminimal gauge breaking there are monopole configurations which
lead to a new interpretation of some of the $4k$ coordinates
parameterizing its moduli space \cite{nonmin}; in these particular
examples there are nonabelian fields present and some parameters
measure the orientation and ``cloud'' size of the nonabelian fields.
Since the low-energy effective $N=2$ theory does not have enhanced
gauge symmetry, these configurations should be suppressed by instanton
effects; our results may shed light on this phenomenon.

Recently, the dimensional reduction of $N=2$ super gauge theories to
$N=4$ supersymmetric gauge theories in $d=2+1$ dimensions has been
investigated in both field theory and string theory
\cite{sw2,ch,dd,hw}.  In this case the (reduced) moduli space of charge $k$
monopoles is isomorphic to the quantum corrected Coulomb branch of
$SU(k)$ gauge theories.  From the gauge theory point of view, the
exponential corrections to the multi-monopole moduli space metric are
obtained through three-dimensional instanton effects. It would be
interesting to see if our results have any implications in this
context.

Finally, a potentially interesting extension of our work comes from
embedding $N=2$ super Yang-Mills theories in string theories and
studying the relation of monopoles to D-branes; to date, such studies
have not focused on the explicit form of quantum corrected stringy
solitons.  Extensions of our work to this context might prove
illuminating both for field theory and for string theory.

\vskip 1em
\noindent{\large\bf Acknowledgements}
\vskip 1em

\noindent{We thank Martin Bucher, Alfred Goldhaber, Amihay Hanany,
Hans Hansson, Ismail Zahed, Philip Argyres, Kimyeong Lee, and Piljin
Yi for discussions.  The work of MR and GC was supported in part by
NSF grant No.~PHY 9309888.}

\appendix
\section{Differential Equation for $u(r)$}

The soliton solutions as derived from the full low-energy theory
are described by the complex function on the order parameter $u(r)$
through the relations on the period integrals (\ref{udependence}).
This gives a complicated transcendental relation between the solitons
in $\f$ space and in $u$-space.  Alternatively one may derive a direct
nonlinear differential equation for $u(r)$.

We consider spherically symmetric dyon configurations of
$\f_D$ and $\f$.  We first re-express the field equations using the
Picard-Fuchs equations \cite{pf}.  For any number of hypermultiplets in
the theory, the periods $a(u)$ and $a_D(u)$ satisfy
\be{PicFuc}
p(u) {d^2 a(u) \over du^2} = - a(u) ,\quad\quad
p(u) {d^2 a_D(u) \over du^2} = - a_D(u)  \ .
\eq
The polynomial $p(u)$ depends in general on $N_f$; for $N_f=0$ it is
determined to be $p(u)=4(u^2-1)$.  Using the spherical ansatz and the
Picard-Fuchs equations, the field equation for the dual scalar
(\ref{dualbox}) may be expressed as
\be{uglytwo}
{\car}e ~e^{i\alpha} \Bigl\{ \partial_u \f_D (r^2 u_r)_r - \f_D \bigl[
  {(r u_r)^2\over p(u)} + 2 L^2 \bigr] \Bigr\} =0  \ ,
\eq
with the same equation for $\f(u)$.

More explicitly, we can substitute the form of $\f(u)$ and $\f_D(u)$ in
terms of complete elliptic integrals given in (\ref{shortint})
to obtain the differential equation
\be{explicitshort}
{\car}e ~{e^{i\alpha} \over (1+u)^{1/2}} \Bigl\{ K(q) (r^2 u_r)_r -
 2 (1+u) E(q) \bigl[ {(r u_r)^2\over p(u)} + 2 L^2 \bigr] \Bigr\} =0 \ .
\eq
The dual scalar field equation is found by substituting
(\ref{longint}) into (\ref{uglytwo}) to obtain
\be{explicitlong}
{\car}e ~{ie^{i\alpha} \over (1+u)^{1/2}} \Biggl\{
 \Bigl(2E(q')-K(q')\Bigr) (r^2 u_r)_r -
 2(1+u) \Bigl( E(q') -K(q')\Bigr) \bigl[ {(r u_r)^2\over p(u)} +
 2 L^2 \bigr] \Biggr\} =0 \ .
\eq
The forms (\ref{explicitshort}) and (\ref{uglytwo}) give
two nonlinear differential equations that determine the
complex order parameter $u(r)$.

\section{$\HH$ Function Expansion}

In this appendix we expand out the contribution from the bosonic portion
of $S_\HH$ in (\ref{N1act}) in terms of $N=0$ components.  We drop the
auxilliary fields and group the terms into the following ten pieces :

\be{A1} 
T_1 &=& {1\over 4} \HH_{AB\bar{C}\bar{D}} \nabla^{\au}\f^{A}
\nabla_{\au}\f^{B}\nabla^{\bu}\bar{\f}^{C}\nabla_{\bu}\bar{\f}^{D}
+\left({1\over 4}\HH_{A\bar{B}\bar{C}} \nabla^{\au}\nabla_{\au}\f^{A}
\nabla^{\bu}\bar{\f}^{B}\nabla_{\bu}\bar{\f}^{C} +c.c.\right)
\nonumber \\ &&
+{1 \over 4} \HH_{A\bar{B}} \nabla^{\au}\nabla_{\au}\f^{A}
\nabla^{\bu}\nabla_{\bu}\bar{\f}^{B}
+{i\over 4}\HH_{A\bar{B}}\nabla^{\a\ad}\f^{A}\left[\nabla_{\ad}^{\b}
f_{\a\b}+\nabla_{\a}^{\bd}f_{\ad\bd},\bar{\f}\right]^{B}
\nonumber \\
T_2 &=& - \HH_{A{\bar B}} f_{CD}^A (\nabla^{\a\bd} f_\bd^\ad)^B \f^C
\nabla_{\a\ad} {\bar\f}^D + \HH_{A{\bar B}{\bar E}} f_{CD}^A \nabla^{\a\bd}
{\bar\f}^E (f_\bd^\ad)^B \f \nabla_{\a\ad} {\bar\f}^D + {\rm c.c.}
\\
T_3 &=& -{1\over 4} \HH_{{\bar A}CD} \nabla_{\au} \nabla^{\au} \f^A
(f^{\a\b})^C (f_{\a\b})^D + {\rm c.c.}
\\
T_4 &=&  \HH_{{\bar A}{\bar B}CD} \nabla^{\b\bd} {\bar\f}^A (f_\bd^\ad)^B
\nabla_\ad^\a \f^C (f_{\a\b})^D
\\
T_5 &=& {1\over 4} \HH_{A{\bar B}} \nabla_{\au} \nabla^{\au} \f^A
\nabla_{\bu} \nabla^{\bu} {\bar\f}^B
\\
T_6 &=& -{1\over 2} f_{AD}^C \HH_{C{\bar B}} \f^D \nabla^{\b\ad} {\bar\f}^B
(\nabla_\b^\bd f_{\bd\ad})^A
\\
T_7 &=& f_{BC}^A f_{AE}^G \HH_{G{\bar D}} \f^E \nabla^{\au} {\bar\f}^D
\f^B \nabla_{\au} {\bar\f}^C
\\
T_8 &=& \HH_{A{\bar B}} (\nabla^{\b\ad} f_\b^\a)^B (\nabla_\a^\md
f_{\md\ad})^A
\\
T_9 &=& {1\over 4} \HH_{A{\bar B}C{\bar D}} (f^{\a\b})^A (f_{\a\b})^C
(f^{\ad\bd})^B (f_{\ad\bd})^D
\\
T_{10} &=& \HH_{{\bar B}CD} (\nabla^{\b\bd} f_\bd^\ad)^B \nabla_\ad^\a
\f^C (f_{\a\b})^D + c.c. \label{A10}
\eq

The energy associated to the above terms may be derived from the
00-component of the energy momentum tensor.  We insert the explicit metric
tensor in each term as in (\ref{general}) and then vary
with respect to them.  The general term in the action contributes
an energy as listed in (\ref{saas}).  We note, however, that 
it is important to track the explicit $\e$ tensors
arising in the bosonic form of $\HH$ since they transform
as tensor densities and thus come with different
powers of $\sqrt{-g}$ when compared with terms only containing the
metric.  Such terms are absent in the case of vanishing electric 
field however, because all covariant time derivatives of the Higgs field
are zero.

As is clear in (\ref{A1}-\ref{A10}) the complete expression for $E$
contains derivatives of $\HH(t)$ with respect to $\phi^{A}$ or
$\bar{\phi}^{A}$.  The potential $\HH$ is only a function of
$t=\frac{\f\cdot\bar\f}{\sqrt{\f^{2}\bar{\f}^{2}}}$, so that we may
find simple expressions for the derivatives.  It is convenient to
express them in terms of the projection operator ${\Pi}_{AB}=
\delta_{AB}-e_A e_B$ on the coset $SU(2)/U(1)$, introduced in
(\ref{CosetAction}), and the unit vector $e^{A}\equiv \langle \f^A
/\sqrt{\f^2} \rangle$.  There are only two generic parameters ${d\over
dt} \HH(t)=\HH'$ and $\HH''$ which in the spherical ansatz are
evaluated at $t=1$.  We give the explicit form of the scalar
derivatives on $\HH$ below:
\be{Hderivatives}
\HH_{A} &\equiv & \frac{\del\HH}{\del\f^{A}} = 0 \nonumber\\
\HH_{A\bar{B}} &=& \frac{\HH\pr}{\f\bar{\f}} {\Pi}_{AB} \nonumber\\
\HH_{AB\bar{C}} &=& -\frac{\HH\pr}{\f^2 \bar{\f}}\left(
  {\Pi}_{AC}e_{B} + {\Pi}_{BC}e_{A}\right)
\\
\HH_{AB\bar{C}\bar{D}} &=& \frac{\HH\prr}{\f^2 \bar{\f}^{2}} \left(
  {\Pi}_{AB}{\Pi}_{CD} + {\Pi}_{AC}{\Pi}_{BD} + 
\Pi_{AD}{\Pi}_{BC} \right)
  \nonumber\\ &&
 +\frac{\HH\pr}{\f^{2}\bar{\f}^{2}} \left({\Pi}_{AB}{\Pi}_{CD} +
  {\Pi}_{AC}e_{B}e_{D} +{\Pi}_{AD}e_{B}e_{C} + 
\Pi_{BC}e_{A}e_{D} + {\Pi}_{BD}e_{A}e_{C}\right)
\nonumber
\eq

Within the radial ansatz the field strengths
decompose in a simple manner onto the $U(1)$ and $SU(2)/U(1)$
directions.  We use the following conventions
\be{fieldstrength}
F_{\mu\nu}^A &=& \partial_\mu A_\nu^A - \partial_\nu A_\mu^A +
\epsilon^A_{~BC} A_\mu^B A_\nu^C \ .
\eq
The electric and magnetic fields, together with the 
covariant derivative on the scalar, are explicitly
\be{FieldDecomps}
E_{i,A} &=& -b_r e_i e_A  -{bL\over r} {\Pi}_{iA}
\\
B_{i,A} &=& -{(1-L^2)\over r^2} e_i e_A  + {L_r\over r} {\Pi}_{iA}
\\
\nabla_i \f_A &=& \f_r e_i e_A  + {L\f\over r} {\Pi}_{iA}  \ .
\eq
Further covariant derivatives of the field strength are not
listed.

The computation of the energy $E$ using the derivatives of 
$H$ in (\ref{Hderivatives})
and the fields (\ref{FieldDecomps}) is straightforward but
intensive.  We first list terms that are proportional to the
constant $\HH\prr(1)$.  The index $i$ in $E_i$ below refers to the term
in (\ref{A1}-\ref{A10}) in which it arises:
\be{dubbel}
E_1\prr &=& \HH\prr \left(\frac{L}{r}\right)^4 \\
E_4\prr &=& \HH\prr \frac{1}{\f\bar{\f}}\left(\frac{L}{r}\right)^2
\left(\left(\frac{bL}{r}\right)^2 - \left(\frac{L_{r}}{r}\right)^2\right) \\
E_{9}\prr &=& \HH\prr
\frac{1}{4\f^2\bar{\f}^2}\left(\left(\frac{L_r}{r}\right)^2
- \left(\frac{bL}{r}\right)^2\right)^2 - \HH\prr
\frac{1}{\f^2\bar{\f}^2} \left(\frac{bL}{r}\right)^4
\eq
The net contribution proportional to $\HH\prr(1)$ may 
then be simplified to
\be{rewrite}
E\prr = \HH\prr \frac{1}{\bigl|\f\bigr|^4}\left(\frac{L}{r}\right)^4 \left\{
 \left(\bigl|\f\bigr|^2 + \half \left( b^2 -
   \left(\frac{L_r}{L}\right)^2 \right)\right)^2 - b^4 \right\} \ .
\eq

The remaining contributing terms to $E$ are proportional to $\HH\pr$.
We suppress the constant $\HH\pr$ and list them below:
\be{sil}
E_1\pr &=& \frac{1}{4} \left\{ \frac{L^{2}(1+L^{2})}{r^4} - \left(\frac{L_{r}}
{r}\right)^{2}-\left(\frac{Lb}{r}\right)^{2} \right\} \\
E_2\pr &=& \left(\frac{L_{r}}{r}\right)^{2} + \left(\frac{bL}
{r}\right)^{2} +\frac{L L_{r}}{2r^{2}}\left( \frac{\f_{r}}{\f}
+\frac{\bar{\f}_{r}}{\bar{\f}}\right) \\
E_4\pr &=& \frac{1}{4\f\bar{\f}}\left\{ \left(\frac{L}{r}\right)^{2}\left[
\frac{(1-L^{2})^{2}}{r^{4}} - 2\left(\frac{L_{r}}{r}\right)^{2}
+2\left(\frac{bL}{r}\right)^{2} - b_{r}^{2}\right]
 \right. \nonumber\\ && \left.
+\Bigl|\frac{\f_{r}}{\f}\Bigr|^{2}
\left(\left(\frac{L_{r}}{r}\right)^{2}-\left(\frac{bL}{r}\right)^2 \right)
\right. \nonumber\\ && \left.
+\left(\frac{L_{r}L(1-L^{2})}{r^{4}}+bb_{r}\left(\frac{L}{r}\right)^{2}
\right)\left(\frac{\f_{r}}{\f}+\frac{\bar{\f}_{r}}  {\bar{\f}}\right)
\right\}
\\
E_6\pr &=& -\frac{Lg}{4r}
\\
E_7\pr &=& \frac{L^2\f\bar{\f}}{r^2}
\\
E_{8}\pr &=& \frac{g^{2}}{4\f\bar{\f}}-
\frac{1}{\f\bar{\f}}\left(\frac{b^{2}L}{r}\right)^{2} \\
E_{9}\pr &=& \frac{1}{2}\left[\frac{L_{r}^{2}-(Lb)^{2}}
{2r^{2}\f\bar{\f}}\right]^2 - \frac{1}{2}\left(\frac{b^{2}L^2}
{r^{2}\f\bar{\f}}\right)^{2}
\\
E_{10}\pr &=& - \frac{g}{4\f\bar{\f}}\left[ 2\frac{L(1-L^{2})}{r^3}
+\frac{L_r}{r}\left(\frac{\f_{r}}{\f}+ \frac{\bar{\f}_{r}}
{\bar{\f}}\right)\right] -\frac{b^3 L^2}{2\f\bar{\f}r^2} i
\left(\frac{\f_{r}}{\f} - \frac{\bar{\f}_{r}}{\bar{\f}}\right) \ .
\eq
We have defined the function $g(r)$ to be
\be{g}
g(r) = \frac{L_{rr}}{r}+\frac{L(1-L^{2})}{r^3}-\frac{Lb^2}{r} \ .
\eq

The net sum from (\ref{sil}) simplifies to the form
\be{Ea}
E_a\pr &=& \HH\pr \frac{1}{2\bigl|\f\bigr|^4}\left(\frac{L}{r}\right)^4 \left\{
 \left(\bigl|\f\bigr|^2 + \half \left( b^2 -
   \left(\frac{L_r}{L}\right)^2 \right)\right)^2 - b^4 \right\}
\\
E_b\pr &=& \HH\pr {1\over 4 r^2 \vert\f\vert^2} \vert L_r {\f_r\over\f}
 + {L(1-L^2)\over r^2} \vert^2
 - \HH\pr {L^2 b^2\over 4r^2\vert\f\vert^2} \vert {b_r\over b}-{\f_r\over\f}
 \vert^2
\\
E_c\pr &=& \HH\pr \left(\frac{L}{2r}-\frac{g}{4\f\bar{\f}}\right) \left\{
2\frac{L(1-L^2)}{r^3}
+\frac{L_r}{r}\left(\frac{\f_r}{\f} + c.c.\right)\right\} \\
E_d\pr &=& \HH\pr\frac{1}{\bigl|\f\bigr|^2}\left\{
\left(\frac{g}{2}-\frac{L\bigl|\f
\bigr|^2}{r}\right)^2-\left(\frac{b^2 L}{r}\right)^2\right\} \label{end}  \ .
\eq
The energy associated to the spherically symmetric soliton
arising from the real nonchiral $\HH$ potential is thus 
given by the sum of (\ref{rewrite}) and (\ref{Ea}-\ref{end}).

\end{document}